\documentclass{iopjournal}

\usepackage{graphicx}% Include figure files
\usepackage{dcolumn}% Align table columns on decimal point
\usepackage{bm}% bold math
\usepackage{tikz}
\usepackage{xcolor}
\begin{document}

\def\gsim{\mathop {\vtop {\ialign {##\crcr 
$\hfil \displaystyle {>}\hfil $\crcr \noalign {\kern1pt \nointerlineskip } 
$\,\sim$ \crcr \noalign {\kern1pt}}}}\limits}
\def\lsim{\mathop {\vtop {\ialign {##\crcr 
$\hfil \displaystyle {<}\hfil $\crcr \noalign {\kern1pt \nointerlineskip } 
$\,\,\sim$ \crcr \noalign {\kern1pt}}}}\limits}

\articletype{Paper} %	 e.g. Paper, Letter, Topical Review...

\title{Ground-state properties of uniaxial magnetic model for antiferromagnetic 1/1 periodic approximants to icosahedral quasicrystals}

\author{Shinji Watanabe
%\orcid
{
0000-0001-9366-3409
}}

\affil{Department of Basic Sciences, Kyushu Institute of Technology, Kitakyushu, Fukuoka 804-8550, Japan}

%\affil{$^2$Department, Institution, City, Country}

%\affil{$^*$Author to whom any correspondence should be addressed.}

\email{swata@mns.kyutech.ac.jp}


\begin{abstract}
The ground-state properties of the uniaxial magnetic model in the 1/1 periodic approximant to the icosahedral quasicrystal with antiferromagnetic interactions for the nearest-neighbor bond $J_1$ and the next-nearest-neighbor bond $J_2$ are analyzed theoretically. 
By performing numerically-exact calculation, we have determined the ground-state phase diagram for the interaction strength $J_2/J_1$ and the magnetic anisotropy defined within the mirror plane reflecting the effect of the crystalline electric field. We have discovered non-collinear magnetic structures, whose magnetic space groups are identified as the antiferromagnetic orders $I_{\rm P}m'\bar{3}'$ and $R_{\rm R}\bar{3}$ with the magnetic propagation vector $(111)$ as well as the antiferromagnetic orders 
%$I222$ 
$Im'm'm'$   
and $R\bar{3}'$ with the magnetic propagation vector $(000)$ and the ferrimagnetic orders $Im'm'2$, $I2'2'2$, $P2'/m'$, and $P\bar{1}$. By analyzing each ground state with degeneracy, we have revealed the magnetic as well as topological properties, which are useful to analyze experimental observations of rare-earth-based 1/1 approximants considering the effect of magnetic domains. 
\end{abstract}

\section{Introduction}

Quasicrystals (QCs) have no three-dimensional periodicity of atoms but posses their structural orders~\cite{Shechtman,Tsai,Takakura}. 
Clarification of the electronic states and their physical properties are frontier of condensed matter physics and material science, 
because the Bloch theorem based on translational symmetry of crystals can no longer be applied. 
One of the fundamental questions about QCs has been whether the magnetic long-range order is realized in QCs~\cite{Fisher1999,Sato2000,Goldman2013,Das2023}. 

There exist periodic crystals which have the common local atomic configuration to those in QCs, which are referred to as approximant crystals (ACs). 
Thus far, experimental efforts have been devoted to seek for the magnetic long-range order in the ACs as well as QCs and  
the ferromagnetic (FM) order and the antiferromagnetic (AFM) order have been observed in the rare-earth based ACs~\cite{Tamura2010,Mori,Hiroto2013,Hiroto2014,Ishikawa2016,Ishikawa2018,Jazbec,Das,Imura,Sawano,Muro,Sato2019,Hiroto,Suzuki2021,Shiino2022,Labib2023,Nawa2023,Labib2024,Gebresenbut,Labib2025,Labib_ACAF_2025}. 
The AFM orders have been observed in the 1/1 AC Cd$_6$R (R=Y, Pr, Nd, Sm, Gd, Tb, Dy, Ho, Er, Tm, Yb, and Lu) ~\cite{Tamura2010,Mori,Das} and also in the 1/1 AC Au-Al-R (R=Gd and Tb)~\cite{Ishikawa2018,Sato2019,Labib2023,Nawa2023,Labib_ACAF_2025}. 
The FM orders have been observed in the 1/1 AC Au-SM-R (SM=Si, Ge, and Al; R=Gd, Tb, Dy, and Ho)~\cite{Hiroto2013,Hiroto2014,Ishikawa2016,Hiroto,Shiino2022,Gebresenbut,Labib2023,Labib2025}.

It is noted that Au-SM-R is the composition tunable compound, which exhibits the FM order and AFM order depending on the composition ratio of Au and SM, e.g., Au$_x$Al$_{86-x}$Gd~\cite{Ishikawa2018}. 
Depending on the compositions, the positive Curie-Weiss temperature and the negative one were observed~\cite{Suzuki2021}. 
This implies that there exist 
%the 
rare-earth-based compounds with 
%the 
FM interactions between the 4f magnetic moments on the rare-earth sites and 
%there 
also 
%exist  
the compounds with 
%the 
AFM interactions.  
Indeed, negative Curie-Weiss temperature was observed, for instance, in the Ce-based 1/1 ACs Ag-In-Ce~\cite{Imura}, Ag-In-(Ce$_x$Y$_{1-x}$)~\cite{Sawano}, and Au-Al-Ce~\cite{Muro}.  
The rare-earth-based ACs with positive and negative Curie-Weiss temperatures are summarized in ref.~\cite{Suzuki2021}. 

Recently, the FM long-range order has been discovered in the QC Au-Ga-R (R=Tb, Gd~\cite{Tamura2021} and Dy~\cite{Takeuchi2023}), which have attracted great interest.  
Moreover, the AFM long-range order has also been discovered in the QC Au-In-Eu~\cite{Tamura2025}, which has attracted much attention. 
Recent theoretical studies on ACs~\cite{Sugimoto2016,Miyazaki2020,WK2021,WPNAS,WSR,WI2023,Sugimoto2024,WFM2024} and QCs~\cite{WPNAS,WSR,WYTF2025,WYTF2026} have also attracted growing interest.

\begin{figure}[bt]
\includegraphics[width=8cm]{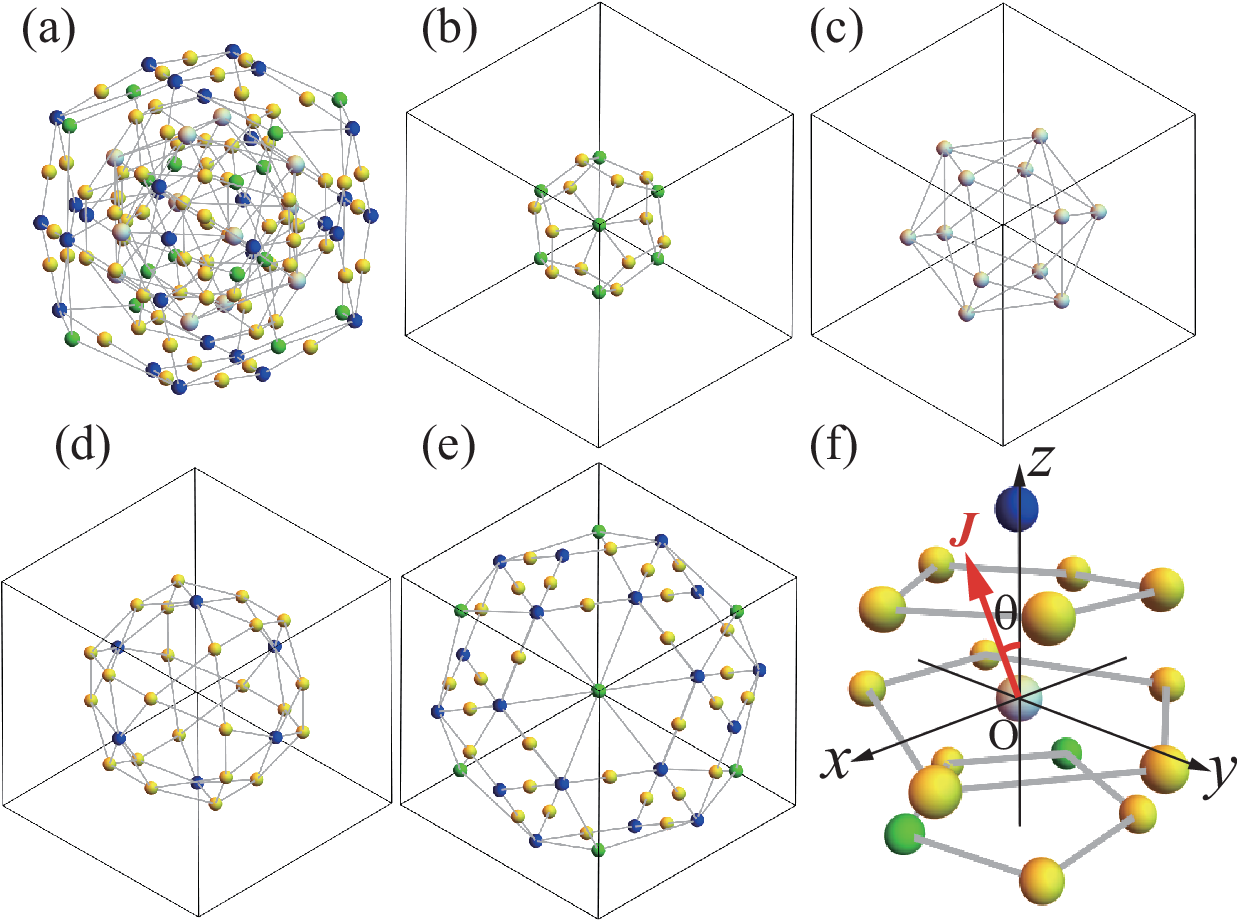}
\caption{
(a) Tsai-type cluster of Au-SM-Ce consists of (b) dodecahedron, (c) icosahedron, (d) icosidodecahedron, and (e) defect rhombic triacontahedron with Au (yellow), SM (blue), Au/SM (green), and Ce (gray). 
The frame box is the unit cell of the (expanded) unit cell of the 1/1 AC Au-SM-Ce. 
(f) The local configuration of atoms surrounding the Ce atom in the Tsai-type cluster. The $z$ axis is the pseudo 5-fold axis. The $yz$ plane is the mirror plane.
The angle $\theta$ is defined between the magnetic easy axis ${\bm J}$ for the CEF ground state and the pseudo 5-fold axis ($z$ axis) within the mirror $(yz)$ plane. 
}
\label{fig:Tb_local}
\end{figure}

These rare-earth based QCs and ACs are composed of the shell structure of atoms referred to as (a) the Tsai-type cluster, which consists of (b) the dodecahedron, (c) the icosahedron, (d) the icosidodecahedron, and (e) the defect rhombic triacontahedron shown in Fig.~\ref{fig:Tb_local}. 
Here we illustrate the atomic coordinates of the 1/1 AC Au$_{65}$Sn$_{20}$Ce$_{15}$ (space group No. 204 $Im\bar{3}$) as a prototypical case~\cite{Kenzari}. 
The frame box of each figure denotes the unit cell of the body-center cubic (bcc) lattice of the 1/1 AC. 
The rare-earth atom is located at the 12 vertices of the icosahedron shown in Fig.~\ref{fig:Tb_local}(c). The 4f electrons at the rare-earth atom are responsible for the magnetism. 

The local configuration of atoms surrounding the rare-earth atom is shown in Fig.~\ref{fig:Tb_local}(f). 
The $z$ axis is the vector passing through the rare-earth atom starting from the center of the icosahedron [see Fig.~\ref{fig:Tb_local}(c)], which is the (pseudo) 5-fold axis in the QC (AC). 
Namely, as seen in Fig.~\ref{fig:Tb_local}(f), the 5-fold configuration of atoms exists around the rare-earth atom. 
Since the 5-fold symmetry is not compatible with translational symmetry of crystals, the well established theory of crystalline electric field (CEF) based on the crystallographic point group cannot be applied to the present system. The lack of the CEF theory has prevented us from understanding the strongly-correlated-electron states in the QCs and ACs, in particular, the magnetic properties~\cite{NKSato2022}. 

%A phenomenological model calculation was performed for the 1/1 AC by assuming the magnetic anisotropy arising from the CEF~\cite{Sugimoto2016}. As for the Gd- or Eu-based systems with Gd$^{3+}$ or Eu$^{2+}$ which have the $4f^7$ configuration with the ground multiplet of $^8S_{7/2}$, the RKKY model was analyzed for the 1/1 AC~\cite{Miyazaki2020,Sugimoto2024}. Recently, Monte Carlo simulation for the Heisenberg model with FM interactions on the icosahedral QC has been performed~\cite{WYTF2025}. 

Recently, 
%the 
general formulation of 
%the 
CEF in the icosahedral QCs and ACs has been developed on the basis of the point charge model~\cite{WK2021}. 
Theoretical analyses of the CEF of the 4f states in the Tsai-type cluster of the Tb-based QC and AC~\cite{WPNAS,WSR} and the Dy-based QC and AC~\cite{WI2023} have revealed that the magnetic easy axis is lying in the mirror plane in the local coordinate
[see Fig.~\ref{fig:Tb_local}(f)]. Namely, the magnetic easy axis for the CEF ground state denoted by ${\bm J}$ is lying in the $yz$ plane, which is the mirror plane.  
The anisotropy angle $\theta$ is defined as the angle between the magnetic easy axis ${\bm J}$ and the pseudo 5-fold axis, i.e., the $z$ axis as Fig.~\ref{fig:Tb_local}(f). 
 
%%%%% Fig.1 %%%%%%%%%%%%%%%%%%%%%%%%%%%%%%%%%%%%%%%%%%%%%%%
%%%%%%%%%%%%%%%%%%%%%%%%%%%%%%%%%%%%%%%%%%%%%%%%%%%%%%%%%

By taking into account the magnetic anisotropy arising from the CEF~\cite{WPNAS}, 
theoretical calculation for the magnetic model on the icosahedron with the FM interaction was performed and the magnetic ground state was obtained~\cite{WPNAS}. Then, on the basis of the magnetic state on the icosahedron, the stable ground state in the 1/1 AC was discussed~\cite{WPNAS}. 
Recently, numerically-exact calculation has been performed on the basis of the magnetic model in the 1/1 AC taking into account the magnetic anisotropy and the ground-state phase diagram for the FM interactions has been determined~\cite{WFM2024}. 
This work has succeeded in explaining the magnetic structures ever identified in the rare-earth-based 1/1 ACs by neutron measurements~\cite{WFM2024}, all of which exhibit positive Curie-Weiss temperatures. However, the magnetic structures of the rare-earth based 1/1 ACs with negative Curie-Weiss temperatures  were not identified experimentally yet.

Theoretically, as for the case of the AFM interactions, single-icosahedron-based analysis was performed~\cite{WSR}. It is interesting to clarify what types of magnetic ground states are realized in the 1/1 AC where the triangular geometry of the icosahedron [see Fig.~\ref{fig:Tb_local}(c)] causes severe frustration due to AFM interactions. In the present study, by performing complete enumeration of all configurations of magnetic moments, we have determined the ground-state phase diagram of the uniaxial model taking account of the CEF with the AFM interactions in the 1/1 AC. Depending on the uniaxial anisotropy and the ratio of the nearest and next-nearest neighbor interactions, eight kinds of noncollinear FM and AFM orders are shown to appear, magnetic space group of each of which is identified. We have revealed the magnetic and topological properties of each magnetic phase.

The organization of this paper is as follows. In Sect.~\ref{sec:MD}, the magnetic model and the method of the numerical calculation is explained. 
The results are presented and their magnetic as well as topological properties are analyzed in Sect.~\ref{sec:RS}.
The topological charges of the whirling state and the vortex state are discussed in Sect.~\ref{sec:TC_IC}. 
The summary and discussion are given in Sect.~\ref{sec:SD}.

%----------------------------------------
\section{Model}
\label{sec:MD}

The rare-earth based 1/1 ACs and QCs are metallic compounds and hence the magnetic interaction mediated by conduction electrons, i.e., the RKKY interaction is considered to play a key role for the quadratic interaction. Moreover, there can exist higher order interactions such as quartic term generally in metallic magnets. Keeping these grounds in our mind, 
we consider the effective model for magnetism taking into account the magnetic anisotropy arising from the CEF in the Au-SM-R~(R=Tb~\cite{WPNAS}, Dy~\cite{WI2023}, and Ce~\cite{WFK2026}). 
\begin{eqnarray}
H=-\sum_{\langle i,j\rangle}J_{ij}\hat{\bm J}_i\cdot\hat{\bm J}_j
\label{eq:H}
\end{eqnarray}
where $J_{ij}$ is the interaction between the magnetic moments at the $i$th site and $j$th site 
and  $\hat{\bm J}_i$ is the unit vector operator parallel or anti-parallel to the magnetic easy axis~\cite{WPNAS,WSR}. 
Namely, the magnetic easy axis lies in the mirror $(yz)$ plane in Fig.~\ref{fig:Tb_local}(f), where the magnetic anisotropy angle $\theta$ is defined by the angle between the magnetic easy axis and the pseudo 5-fold $(z)$ axis as shown in Fig.~\ref{fig:Tb_local}(f).
%Recently, 
As noted above, 
this model with the FM interactions has been studied theoretically and the ground-state phase diagram has been determined by numerically-exact calculations~\cite{WFM2024}, where 
the measured magnetic structures of the AFM as well as FM long-range orders in the Tb- and Ho-based 1/1 ACs~\cite{Sato2019,Hiroto,Labib2023,Gebresenbut} are explained. 
Hence, we consider that this model captures the essence of the magnetism of the rare-earth based ACs.  

\begin{figure}[tb]
\includegraphics[width=7cm]{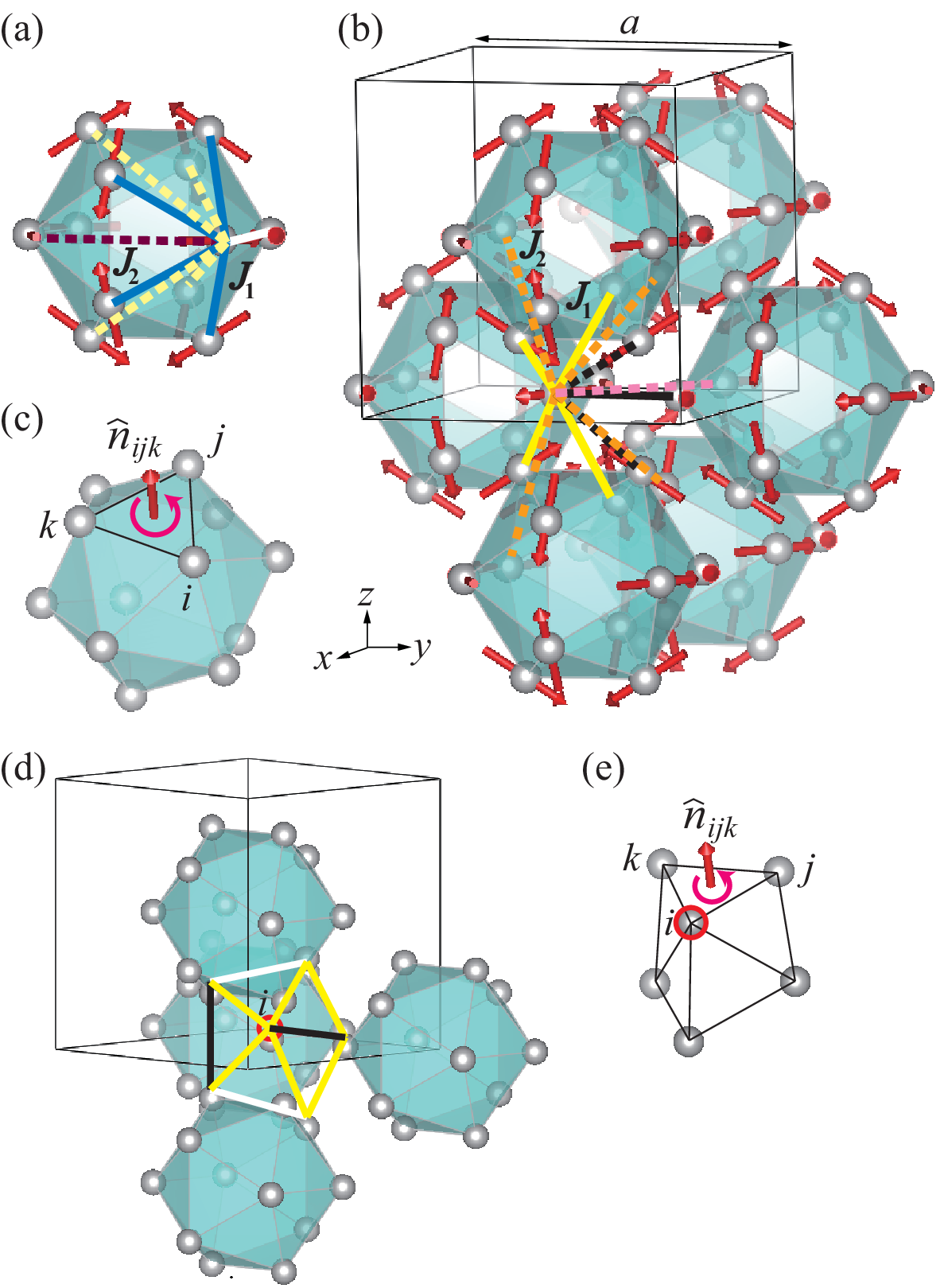}%
\caption{
The nearest-neighbor (N.N.) interactions $J_1$ (solid line) and next-nearest-neighbor (N.N.N.) interactions $J_2$ (dashed line) for (a) the intra icosahedron and (b) the inter icosahedron are illustrated. 
In (b),  the unit cell of the bcc lattice of the 1/1 AC is denoted with the lattice constant $a$. 
Unit normal vector $\hat{n}_{ijk}$ (red arrow) of the triangular surface formed by the $i$, $j$, and $k$ sites defined by the anticlockwise direction for (c) icosahedron and (e) pentagonal pyramid. (d) Pentagonal pyramid is composed of inter-icosahedron vertices. 
}
\label{fig:lattice}
\end{figure}

As for the lattice structure of the rare-earth sites of the 1/1 AC [see Fig.~\ref{fig:Tb_local}(c)], following ref.~\cite{WFM2024},   
we employ the atomic coordinates identified by the X-ray measurement in the 1/1 AC Au$_{70}$Si$_{17}$Tb$_{13}$ with a lattice constant $a=14.726~{\rm \AA}$ as a prototypical system~\cite{Hiroto}. 
The interaction $J_{ij}$ is set as follows~\cite{WFM2024}: 
As for intra icosahedron interactions, 
$J_1$ is set for the 5 bonds for each Tb site with the bond length being $0.374a$ (1 bond represented by the white line) and $0.378a$ (4 bonds  represented by blue lines) and $J_2$ is set for the 5 bonds for each Tb site with the bond length being $0.610a$ (4 bonds  represented by yelllow dashed lines) and $0.612a$ (1 bond  represented by the purple dashed line) [see Fig.~\ref{fig:lattice}(a)]. 
%Here, $a$ is the lattice constant of the 1/1 AC. 
As for inter icosahedron interactions, $J_1$ is set for the 5 bonds for each Tb site with the bond length being $0.368a$ (4 bonds represented by yellow lines) and $0.388a$ (1 bond represented by  the black line) and $J_2$ is set for the 7 bonds with the bond length being $0.528a$ (2 bonds represented by black dashed lines), $0.530a$ (4 bonds represented by orange dashed lines), and $0.539a$ (1 bond  represented by the pink dashed line)  [see Fig.~\ref{fig:lattice}(b)].

%%%%%%%% Fig.2 %%%%%%%%%%%%%%%%%%%%%%%%%%%%%%%
%%%%%%%%%%%%%%%%%%%%%%%%%%%%%%%%%%%%%%%%%%%%

In this study, we analyze  the model (\ref{eq:H}) in the 1/1 AC with AFM interactions, i.e., $J_1<0$ and $J_2<0$. 
The unit cell considered in this study is displayed as the frame box in Fig.~\ref{fig:lattice} where the two icosahedrons with 24 Tb sites are contained. 
We have calculated the 
%expectation 
eigenvalues of the energy of the model (\ref{eq:H}) under the periodic boundary condition 
%by all the eigenstates 
and have obtained the ground state(s) with the lowest energy. 

In ref.~\cite{WPNAS,WSR}, the topological charge of the magnetic texture on the icosahedron was introduced.   
In this study, we have calculated the topological charge defined as 
the solid angle divided by the surface area of the unit sphere: 
\begin{eqnarray}
n=\frac{\Omega}{4\pi},
\label{eq:tp}
\end{eqnarray}
where $\Omega$ is the solid angle subtended by the twelve magnetic moments on the icosahedron 
\begin{eqnarray}
\Omega=\sum_{i,j,k\in {\rm IC}}\Omega_{ijk}.
\label{eq:SI_IC}
\end{eqnarray}
Here, $\Omega_{ijk}$ is the solid angle subtended by the three magnetic moments at the $i$, $j$, and $k$ sites on each triangular surface of the icosahedron,
\begin{eqnarray}
\Omega_{ijk}=2\tan^{-1}\left[\frac{\chi_{ijk}}{1+{\bm J}_i\cdot{\bm J}_j+{\bm J}_j\cdot{\bm J}_k+{\bm J}_k\cdot{\bm J}_i}\right]
\label{eq:SI_tri}
\end{eqnarray}
\cite{Eriksson}, where ${\bm J}_i$ is the magnetic moment at the $i$th site and $\chi_{ijk}$ is the scalar chirality defined by
\begin{eqnarray}
\chi_{ijk}={\bm J}_i\cdot({\bm J}_j\times{\bm J}_k). 
\label{eq:SC}
\end{eqnarray}
%

%The scalar chirality of the icosahedron with its center-of-mass position ${\bm R}$ is defined by
%%
%\begin{eqnarray}
%\chi({\bm R})=\sum_{i,j,k\in{\rm IC}}\chi_{ijk},
%\label{eq:SC2}
%\end{eqnarray}
%%
%the order of $i$, $j$, and $k$ is defined in the anticlockwise direction with respect to the normal vector of the triangle formed by the $i$th, $j$th, and $k$th sites, $\hat{\bm n}_{i,j,k}$, pointing outward from ${\bm R}$. 

We have also calculated the total chirality defined by
\begin{eqnarray}
{\bm \chi}^{\rm T}=\sum_{\langle i,j,k\rangle}\chi_{ijk}\hat{n}_{ijk},
\label{eq:chi_T}
\end{eqnarray}
where the summation is taken over all the three sites on the icosahedron and $\hat{n}_{ijk}$ denotes the surface normal~\cite{Aoyama2021}. 
The total chirality plays a role as the emergent fictitious magnetic field, which gives rise to the topological Hall effect via the relation~\cite{Tatara}
\begin{eqnarray}
\sigma_{\mu\nu}^{\rm T}\propto\epsilon_{\mu\nu\rho}\chi_{\rho}^{\rm T}.
\label{eq:sigma}
\end{eqnarray}
Here, 
%where
 $\epsilon_{\mu\nu\rho}$ is the Levi-Civita symbol defined by  
\begin{eqnarray}
\epsilon_{\mu\nu\rho}=
\left\{
\begin{array}{rl}
1, & (\mu,\nu,\rho)=(1,2,3), (2,3,1), (3,1,2) \\
-1, & (\mu,\nu,\rho)=(2,1,3), (1,3,2), (3,2,1) \\
0, & {\rm otherwise}. 
\end{array}
\right.
\label{eq:LC}
\end{eqnarray}

Since it was shown in Ref.~\cite{Tatara} that exponentially decaying factor appears in the proportionality constant in Eq.~(\ref{eq:sigma}), we consider the triads with the short side lengths denoted by solid lines in Figs.~\ref{fig:lattice}(a) and \ref{fig:lattice}(b) with the lengths being $0.368a\sim 0.388a$. Hence, we calculated $\chi_{ijk}$ on the 20 triangular surfaces of the icosahedron [see Fig.~\ref{fig:lattice}(c)] and 5 triangles for inter icosahedron [see Fig.~\ref{fig:lattice}(d)]. Here, $\chi_{ijk}$ in Eq.~(\ref{eq:SC}) is calculated with the order of $i$, $j$, and $k$ being defined in the anticlockwise direction with respect to the unit normal vector $\hat{n}_{ijk}$ pointing outward from the center of the icosahedron and the pentagonal pyramid shown in Figs.~\ref{fig:lattice}(c) and \ref{fig:lattice}(e), respectively. In this study, we show the result of ${\bm \chi}^{\rm T}$ per icosahedron calculated in this way by Eq. (\ref{eq:chi_T}).

We have also calculated the total magnetic moment per icosahedron defined by
\begin{eqnarray}
{\bm J}_{\rm IC}=\sum_{i=1}^{12}{\bm J}_{i}. 
\label{eq:J_IC}
\end{eqnarray}

We have performed the numerically-exact calculation of the model (\ref{eq:H}) by inputting various $J_2 (<0)$ and $J_1 (<0)$ with the anisotropy angle $\theta$. 
We have calculated all the energies of all the eigenstates and have identified the ground states which have the lowest energy with the degeneracy. It is noted that our calculation is equivalent to the exact diagonalization method. 
For the obtained ground state, we have calculated the topological charge $n$ in Eq.~(\ref{eq:tp}), total chirality ${\bm \chi}^{\rm T}$ in Eq.~(\ref{eq:chi_T}), and the total magnetic moment per the icosahedron ${\bm J}_{\rm IC}$ in Eq.~(\ref{eq:J_IC}). 

%---------------------------------------------------
\section{Results}
\label{sec:RS}

\subsection{Ground state phase diagram}

\begin{figure}[tb]
\includegraphics[width=8cm]{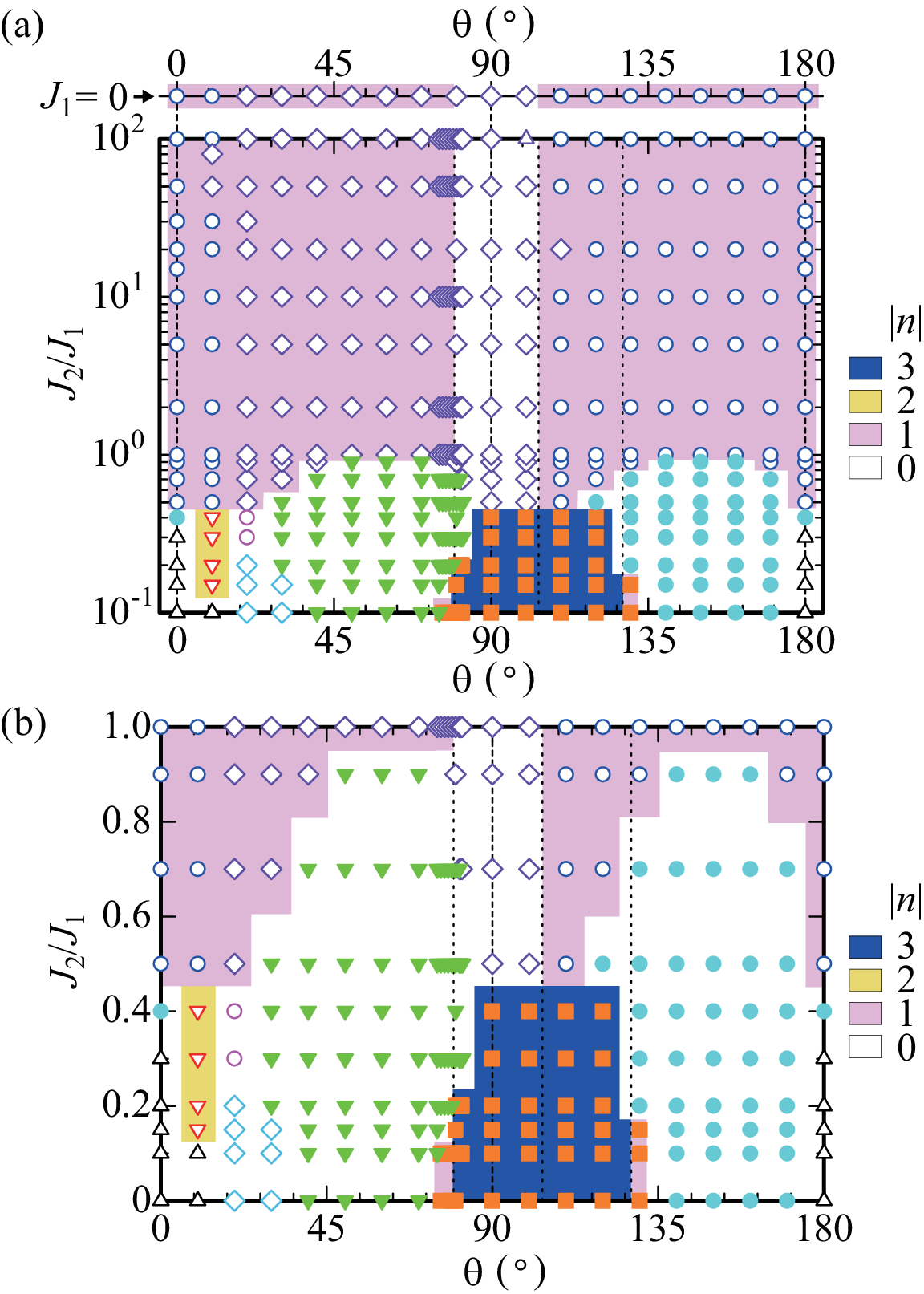}
\caption{
The ground-state phase diagram in the plane of $J_2/J_1$ and the anisotropy angle $\theta$. 
(a) $J_1=0$ (upper panel) and $10^{-1}\le J_2/J_1\le 10^2$ in logarithmic plot (lower panel). 
(b) The enlargement for $0\le J_2/J_1\le 1$. 
In (a) and (b), the colored regions denote magnetic orders characterized by the topological charge $|n|=3$ (blue), $2$ (yellow), $1$ (pink), and $0$ (white). 
The open symbols denote uniform orders and the filled symbols denote antiferromagnetic orders. 
%The gray symbols denote the ordered states where the magnetic states on the icosahedrons located at the center and corner of the bcc unit cell are different.
The vertical dashed lines at $\theta=0^{\circ}$, $90^{\circ}$ and $180^{\circ}$ are the guides for the eyes. The vertical dotted lines at $\theta=\theta_{\rm c1}$, $\theta_{\rm c2}$ and $\theta_{\rm c3}$ are the guides for the eyes (see text).  
}
\label{fig:PD}
\end{figure}

The ground-state phase diagram in the plane of $J_2/J_1$ and the anisotropy angle $\theta$ is shown in Fig.~\ref{fig:PD}. 
The result shows that non-collinear and non-coplanar magnetic structures are stabilized depending on the ratio of interactions $J_2/J_1$ and the anisotropy angle $\theta$. 
In Fig.~\ref{fig:PD}(a), $10^{-1}\le J_2/J_1\le 10^{2}$ in logarithmic scale vs $\theta$ as well as the phase diagram for $J_1=0$ is shown. In Fig.~\ref{fig:PD}(b), the enlargement for $0\le J_2/J_1\le 1$ vs $\theta$ is shown. 
As noted above Eq. (1), the RKKY interaction is considered to contribute to the interaction of the effective model Eq. (1). Since the RKKY interaction has the oscillatory nature, the magnitude of the interaction for longer distances can be larger than the one for shorter distances depending on the density of the conduction electrons. At the node of the RKKY interaction, the interaction can be zero. Hence, we have determined the phase diagram for $0\le J_2/J_1\le 10^2$ including both limits of $J_1=0$ and $J_2=0$. 

The open symbols denote the magnetic states, which are uniformly distributed at the center and corner icosahedrons of the bcc unit cell, i.e., the uniform states. Hence, these states are specified by the magnetic translation vector $(hkl)=(000)$ in ${\bm q}=\frac{2\pi}{a}(hkl)$ with $a$ being the lattice constant. 
The filled symbols with color denote the magnetic states, where the magnetic moments at the corner icosahedrons are obtained by inverting all the magnetic moments at the center icosahedron in the bcc unit cell. Hence, these states are specified by the magnetic translation vector $(hkl)=(111)$, which are referred to as the antiferromagnetic states. 
Since the time-reversal state, i.e., the magnetic state with all the magnetic moments being inverted, is energetically degenerate, the same ground state at $\theta=0^{\circ}$ appears at $\theta=180^{\circ}$. 
The magnetic orders characterized by the topological charge $|n|$ are denoted by the colored regions for $|n|=3$ (blue), $2$ (yellow), $1$ (pink), and $0$ (white). 
Below the property of each magnetic state for the antiferromagnetic orders and for uniform orders is explained.

%%%%% Fig.3 %%%%%%%%%%%%%%%%%%%%%%%%%%%%%%%%%%%%%%%%%%%%%%%
%%%%%%%%%%%%%%%%%%%%%%%%%%%%%%%%%%%%%%%%%%%%%%%%%%%%%%%%%

%-------------------------------------------------
\subsubsection{Antiferromagnetic orders}

\begin{figure}[tb]
\includegraphics[width=12cm]{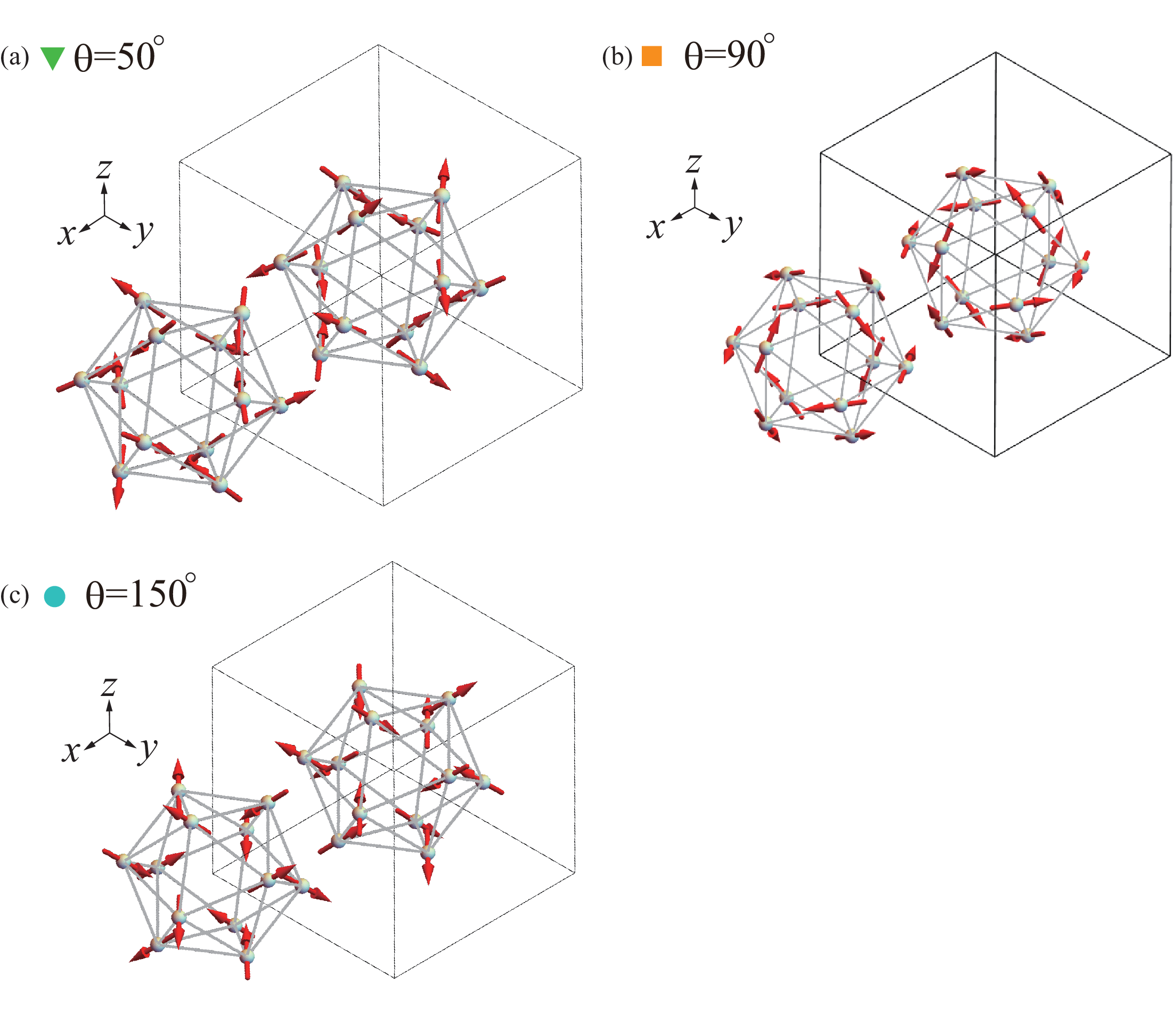}
\caption{(a) 
Magnetic state with anisotropy angle  
$\theta=50^{\circ}$ viewed from the three-fold axis (111) direction.
(b) Whirling-anti-whirling state with anisotropy angle $\theta=90^{\circ}$ viewed from the one of the 3-fold-axis directions. 
(c) Magnetic state with anisotropy angle $\theta=150^{\circ}$ viewed from the three-fold axis (111) direction. 
}
%\label{fig:grn_inv_tri}
\label{fig:AF}
\end{figure}

The magnetic order denoted by the filled green inverted triangle is stabilized for $30^{\circ}\lsim\theta\lsim 90^{\circ}$ and small $J_2/J_1$ in the phase diagram Fig.~\ref{fig:PD}. 
The magnetic space group is $R_{\rm R}\bar{3}$ (No. 148.4.1250 in the OG notation~\cite{OG} and No. 148.20 $RI\bar{3}$ in the BNS notation~\cite{BNS}). 
From the space group $Im\bar{3}$ to rhombohedral $R_{\rm R}\bar{3}$, one 3-fold axis is singled out as a unique axis and it turns out that total magnetic-moment vector per icosahedron is directed along the unique axis. Namely, the unique axis of the magnetic state shown in 
%Fig.~\ref{fig:grn_inv_tri} 
Fig.~\ref{fig:AF}(a)  
is (111) where we have obtained the total magnetic moment ${\bm J}_{\rm IC}=(-0.58,-0.58,-0.58)$ and total chirality ${\bm \chi}^{\rm T}=(-0.06,-0.06,-0.06)$ per icosahedron located at the corner of the frame box in 
%Fig.~\ref{fig:grn_inv_tri}. 
Fig.~\ref{fig:AF}(a). 
Note that all the magnetic moments at the center icosahedron are directed oppositely, giving $-{\bm J}_{\rm IC}$ and  $-{\bm \chi}^{\rm T}$ so that the net magnetic moment and total chirality are cancelled in the unit cell. The topological charge is zero $n=0$ for both states at corner and center icosahedrons in 
%Fig.~\ref{fig:grn_inv_tri}. 
Fig.~\ref{fig:AF}(a). 
The way of choosing one 3-fold axis among four diagonal directions of the cube is four. These four degrees with time-reversal degree give rise to the eight domains. Our numerically-exact calculation has confirmed this as the degeneracy of this ground state is 8 for $J_2<0$, i.e., $4\times 2=8$. These characteristics are summarized in Table~\ref{tb:mag_state}.  At the 5th, 6th and 7th column, ${\bm J}_{\rm IC}$, ${\bm \chi}^{\rm T}$, and $n$ for each degenerate state are listed, respectively. Note that each time-reversal state with the same ground-state energy has $-{\bm J}_{\rm IC}$, $-{\bm \chi}^{\rm T}$, and $-n$, which are not listed in the  5th, 6th and 7th column in Table~\ref{tb:mag_state}. We note that the degeneracy for $J_2=0$ is 64, where the magnetic state shown in 
%Fig.~\ref{fig:grn_inv_tri} 
Fig.~\ref{fig:AF}(a) 
is included in the degenerated states. 

%%%%% Fig.4 %%%%%%%%%%%%%%%%%%%%%%%%%%%%%%%%%%%%%%%%%%%%%%%
%%%%%%%%%%%%%%%%%%%%%%%%%%%%%%%%%%%%%%%%%%%%%%%%%%%%%%%%%

\begin{figure}[tb]
\includegraphics[width=8cm]{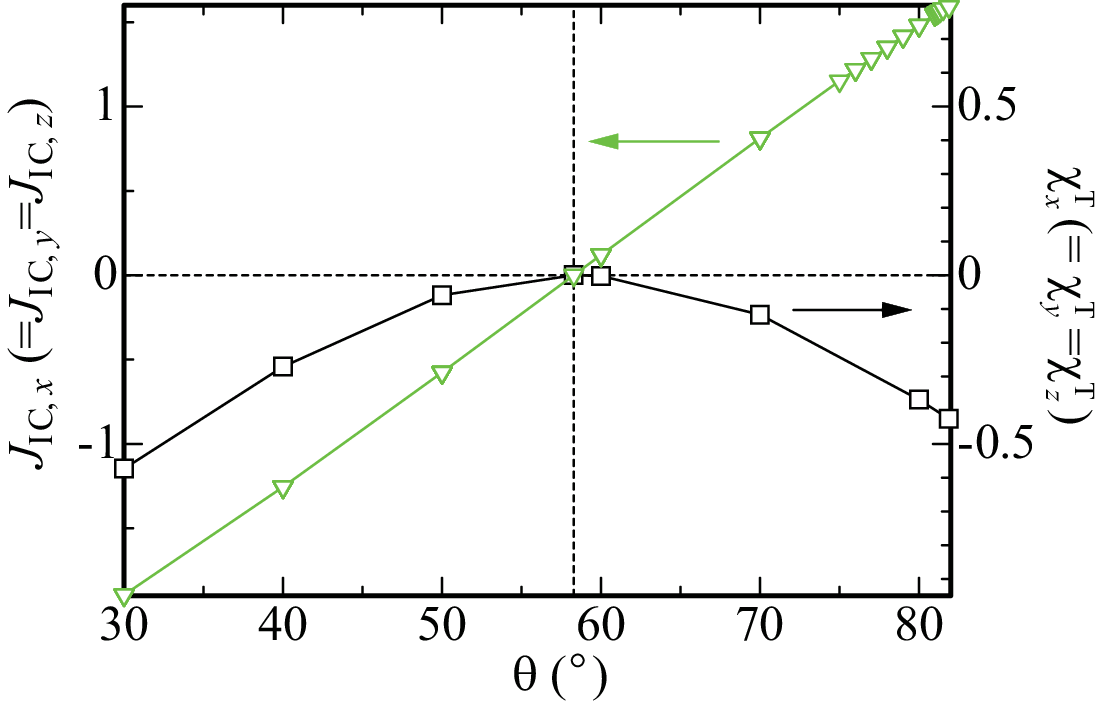}
\caption{The anisotropy-angle dependence of each component of the total magnetic moment per icosahedron $J_{{\rm IC},x}$ and the total chirality ${\chi}^{\rm T}_{x}$ for the magnetic state denoted by filled green inverted triangle in Fig.~\ref{fig:PD}. The vertical dashed line at $\theta=\theta_0=58.2825\dots^{\circ}$ is the guide for the eyes. }
\label{fig:gr_inv_tri_th}
\end{figure}

Figure~\ref{fig:gr_inv_tri_th} shows the anisotropy-angle dependence of $x$ component of the total magnetic moment per icosahedron $J_{{\rm IC},x}$ and the total chirality ${\chi}^{\rm T}_{x}$ for the magnetic state denoted by the filled green inverted triangle symbol in Fig.~\ref{fig:PD}. Here we show the result for the state listed at the first row in Table~\ref{tb:mag_state} with $J_{{\rm IC},x}=J_{{\rm IC},y}=J_{{\rm IC},z}$ and ${\chi}^{\rm T}_{x}=\chi^{\rm T}_{y}=\chi^{\rm T}_{z}$. As $\theta$ increases, $J_{{\rm IC},\mu}$ $(\mu=x, y, z)$ monotonically  increases with the sign being changed at $\theta=\theta_0\equiv\tan^{-1}(\tau)=58.28254\dots^{\circ}$~\cite{WFM2024}, where $\tau$ is the golden mean $\tau=(1+\sqrt{5})/2$.  The total chirality $\chi^{\rm T}_{\mu}$ $(\mu=x, y, z)$ does not change the sign while the absolute value decreases toward $\theta_0$.

%%%%% Fig.5 %%%%%%%%%%%%%%%%%%%%%%%%%%%%%%%%%%%%%%%%%%%%%%%
%%%%%%%%%%%%%%%%%%%%%%%%%%%%%%%%%%%%%%%%%%%%%%%%%%%%%%%%%

The whirling-anti-whirling order is realized in the region of $90^{\circ}\le\theta\le130^{\circ}$ for small $J_2/J_1$ in the phase diagram denoted as filled orange square symbols shown in Fig.~\ref{fig:PD}. 
The magnetic space group of the whirling-anti-whirling state is $I_{\rm P}m'\bar{3'}$ (No. 204.5.1534 in the OG notation~\cite{OG} and No. 201.21 $Pn\bar{3}$ in the BNS notation~\cite{BNS})~\cite{WFM2024}.
The view of this order at $\theta=90^{\circ}$ from one of the three-fold axes, i.e., the (111) direction, is plotted in 
%Fig.~\ref{fig:or_sq}. 
Fig.~\ref{fig:AF}(b) 
The total magnetic moment on the icosahedron is zero ${\bm J}_{\rm IC}={\bf 0}$ and the total chirality is zero ${\bm \chi}^{\rm T}={\bf 0}$. 
The whirling-moment state on the icosahedron at the center of the unit cell is characterized by the topological charge $n=+3$ and the anti-whirling-moment state on the icosahedron at the corner of the unit cell is characterized by the topological charge $n=-3$~\cite{WPNAS}. Hence, this state is the topological magnetic texture characterized by finite topological charge. 
This magnetic state preserves cubic symmetry, i.e., high symmetry is kept from $Im\bar{3}$ to  $I_{\rm P}m'\bar{3'}$.  
The magnetic state where all the magnetic moments are inverted is also energetically degenerate with the state shown in 
%Fig.~\ref{fig:or_sq} 
Fig.~\ref{fig:AF}(b) 
so that the degeneracy of the ground state is 2. 
These characteristics are summarized in Table~\ref{tb:mag_state}.

%%%%% Fig.6 %%%%%%%%%%%%%%%%%%%%%%%%%%%%%%%%%%%%%%%%%%%%%%%
%\begin{figure}[tb]
%\includegraphics[width=8cm]{or_sq_AC_(111)_th_90.eps}
%\caption{Whirling-anti-whirling state with anisotropy angle $\theta=90^{\circ}$ viewed from the one of the 3-fold-axis directions.
%}
%\label{fig:or_sq}
%\end{figure}
%%%%%%%%%%%%%%%%%%%%%%%%%%%%%%%%%%%%%%%%%%%%%%%%%%%%%%%%%

It is noted here that the  whirling-anti-whirling order was observed in the 1/1 AC Au-Al-Tb with positive Curie-Weiss temperature~\cite{Sato2019,Nawa2023,Labib2023}. The magnetic structure has been reproduced recently by the numerically-exact calculation on the basis of the effective model~(\ref{eq:H}) with the FM interactions $J_1>0$ and $J_2>0$~\cite{WFM2024}.   
Interestingly, it has been theoretically shown that by applying a magnetic field to the whirling-anti-whirling order, the metamagnetic transition takes place simultaneously with the topological transition~\cite{WPNAS,WFM2024}. In the high-field phase after the metamagnetic transition, it has also been pointed out that the topological Hall effect is expected to appear~\cite{WPNAS,WFM2024}. Recently, the way of controlling the domains has been proposed theoretically. Namely, the magnetic-field-direction dependence on the topological Hall effect from the 2-fold axis (001) to the 3-fold axis (111) has been clarified theoretically~\cite{WICM2024}.

The magnetic state denoted by the filled light-blue circle symbol in Fig.~\ref{fig:PD} is stabilized for $120^{\circ}\lsim\theta\lsim 180^{\circ}$ and small $J_2/J_1$. 
The magnetic space group is $R_{\rm R}\bar{3}$ (No. 148.4.1250 in the OG notation~\cite{OG} and No. 148.20 $RI\bar{3}$ in the BNS notation~\cite{BNS}), which is the same magnetic space group as the magnetic state shown in 
%Fig.~\ref{fig:grn_inv_tri}. 
Fig.~\ref{fig:AF}(a). 
Namely, the unique axis of the magnetic state shown in 
%Fig.~\ref{fig:lbl_circ} 
Fig.~\ref{fig:AF}(c)  
is (111) where we have obtained the total magnetic moment ${\bm J}_{\rm IC}=(0.12,0.12,0.12)$ and total chirality ${\bm \chi}^{\rm T}=(-0.22,-0.22,-0.22)$ per icosahedron located at the corner of the frame box in 
%Fig.~\ref{fig:grn_inv_tri}.   
Fig.~\ref{fig:AF}(a). 
At the center icosahedron, all the magnetic moments are directed oppositely, giving $-{\bm J}_{\rm IC}$ and  $-{\bm \chi}^{\rm T}$ so that the net magnetic moment and total chirality are cancelled in the unit cell. The topological charge is zero $n=0$ for each state located at the corner and center icosahedrons in 
%Fig.~\ref{fig:lbl_circ}. 
Fig.~\ref{fig:AF}(c). 

Our numerically-exact calculation has shown eightfold degeneracy for the ground state for $J_2<0$, which corresponds to the number of domains from $Im\bar{3}$ to $R_{\rm R}\bar{3}$ as noted above. 
These characteristics are summarized in Table~\ref{tb:mag_state}. We note that the degeneracy for $J_2=0$ is 64, where the magnetic state shown in 
%Fig.~\ref{fig:lbl_circ} 
Fig.~\ref{fig:AF}(c)  
is included in the degenerated states.

%%%%% Fig.7 %%%%%%%%%%%%%%%%%%%%%%%%%%%%%%%%%%%%%%%%%%%%%%%
%\begin{figure}[tb]
%\includegraphics[width=8cm]{AF_lbl_circ_AC_(111)_th_150.eps}
%\caption{Magnetic state with anisotropy angle $\theta=150^{\circ}$ viewed from the three-fold axis (111) direction.}
%\label{fig:lbl_circ}
%\end{figure}
%%%%%%%%%%%%%%%%%%%%%%%%%%%%%%%%%%%%%%%%%%%%%%%%%%%%%%%%%

\begin{figure}[tb]
\includegraphics[width=8cm]{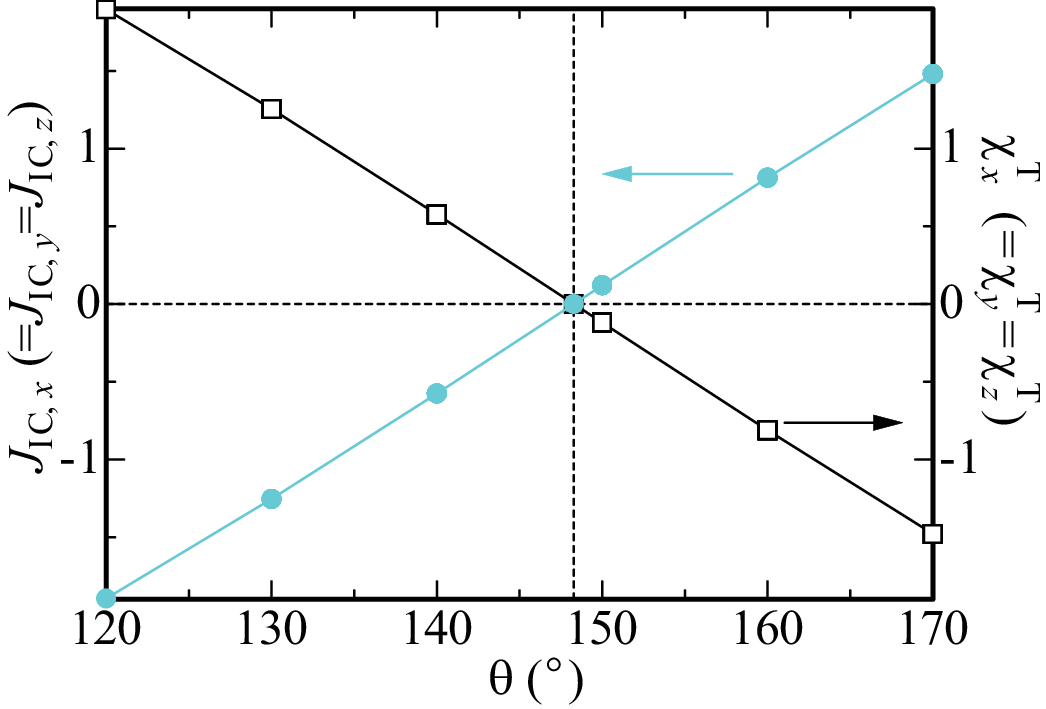}
\caption{
The anisotropy-angle dependence of each component of the total magnetic moment per icosahedron $J_{{\rm IC},x}$ and the total chirality ${\chi}^{\rm T}_{x}$ for the magnetic state denoted by filled light-blue circle triangle in Fig.~\ref{fig:PD}. The vertical dashed line at $\theta=\theta_1=148.2825\dots^{\circ}$ is the guide for the eyes. 
}
\label{fig:lbl_circ_th}
\end{figure}

Figure~\ref{fig:lbl_circ_th} 
shows the anisotropy-angle dependence of each component of the total magnetic moment per icosahedron $J_{{\rm IC},x}$ and the total chirality ${\chi}^{\rm T}_{x}$ for the magnetic state denoted by the filled light-blue circle in Fig.~\ref{fig:PD}. Here we show the result for the state listed at the first row in Table~\ref{tb:mag_state}, where $J_{{\rm IC},x}=J_{{\rm IC},y}=J_{{\rm IC},z}$ and ${\chi}^{\rm T}_{x}=\chi^{\rm T}_{y}=\chi^{\rm T}_{z}$ hold. As $\theta$ increases, $J_{{\rm IC},\mu}$ $(\mu=x, y, z)$ monotonically  increases where the sign changes at $\theta=\theta_1$ with $\theta_1$ being $\theta_1\equiv\theta_0+90^{\circ}=148.28254\dots^{\circ}$~\cite{WFM2024}.  The total chirality $\chi^{\rm T}_{\mu}$ $(\mu=x, y, z)$ monotonically decreases with sign change at $\theta=\theta_1$, as $\theta$ increases.

%%%%% Fig.8 %%%%%%%%%%%%%%%%%%%%%%%%%%%%%%%%%%%%%%%%%%%%%%%
%%%%%%%%%%%%%%%%%%%%%%%%%%%%%%%%%%%%%%%%%%%%%%%%%%%%%%%%%

%%%%% Fig.7 %%%%%%%%%%%%%%%%%%%%%%%%%%%%%%%%%%%%%%%%%%%%%%%
%\begin{figure}[tb]
%\includegraphics[width=8cm]{st.eps}
%\caption{
%Staggered orders  in the 1/1 AC viewed from the three-fold axis (111) direction: 
%(a) the ... order at $\theta=50^{\circ}$, 
%(b) the whirling-anti-whirling order at $\theta=90^{\circ}$
%and (c) the ... order at $\theta=150^{\circ}$.
%}
%\label{fig:st}
%\end{figure}
%%%%%%%%%%%%%%%%%%%%%%%%%%%%%%%%%%%%%%%%%%%%%%%%%%%%%%%%%

%------------------------------------------------------------------

\subsection{Uniform orders}

In this section, we discuss the property of uniform orders specified by the magnetic translation vector ${\bm q}=\frac{2\pi}{a}(hkl)={\bf 0}$ in the phase diagram Fig.~\ref{fig:PD}.
The magnetic states are classified into two classes. One is the magnetic state where the total magnetic moment per icosahedron ${\bm J}_{\rm IC}$ is zero, which we refer to as the $\Gamma$-point antiferromagnetic state. 
The other is nonzero ${\bm J}_{\rm IC}\ne{\bf 0}$, which is referred to as ferrimagnetic state. 
Below we discuss the properties of the magnetic states in each class. 

%-------------------------------------------------------------
\subsubsection{ $\Gamma$-point antiferromagnetic orders}

%The non-collinear magnetic state denoted by the brown triangle is realized for small $\theta$ and large $J_2/J_1$ in the ground-state phase diagram shown in Fig.~\ref{fig:PD}. 
%The magnetic structure for $\theta=10^{\circ}$ is illustrated in Fig.~\ref{fig:br_tri}. 
%The magnetic space group is $I222$ No. 23.1.145 in the OG notation (No. 23.49 in the BNS notation). 
%The total magnetic moment on the icosahedron is zero ${\bm J}_{\rm IC}={\bf 0}$ and the total chirality is zero ${\bm \chi}^{\rm T}={\bf 0}$. 
%%This magnetic state is characterized by the topological charge $n=-1$. 
%%The magnetic state where all the magnetic moments are inverted, which is characterized by the topological charge $n=+1$, is also energetically degenerate with the state shown in Fig.~\ref{fig:br_tri} so that the degeneracy of the ground state is 2. 
%These characteristics are summarized in Table~\ref{tb:mag_state}. 

%%%%% Fig.9 %%%%%%%%%%%%%%%%%%%%%%%%%%%%%%%%%%%%%%%%%%%%%%
%\begin{figure}[bt]
%\includegraphics[width=6cm]{br_tri_IC_th_10.eps}
%\caption{
%Non-collinear state with anisotropy angle $\theta=10^{\circ}$.}
%\label{fig:br_tri}
%\end{figure}
%%%%%%%%%%%%%%%%%%%%%%%%%%%%%%%%%%%%%%%%%%%%%%%%%%%%%%%%%

\begin{figure}[tb]
\includegraphics[width=4.5cm]{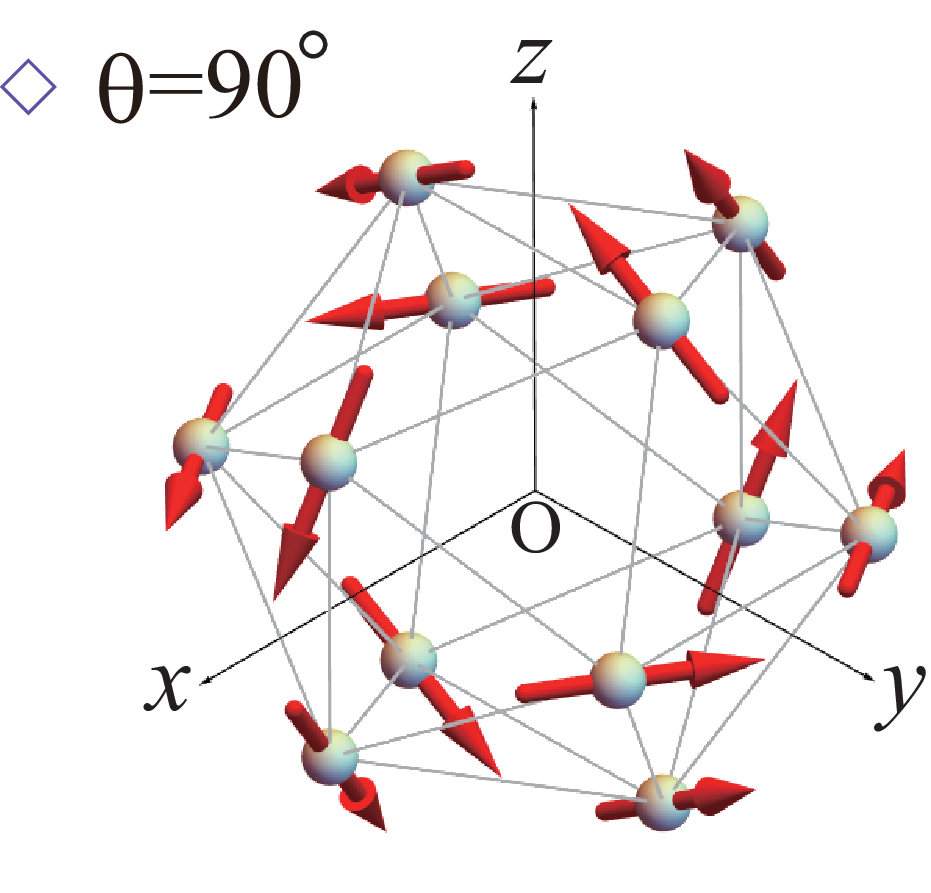}
\caption{
Vortex state with anisotropy angle $\theta=90^{\circ}$ viewed from the (111) direction. 
}
\label{fig:prp_dia}
\end{figure}

The magnetic state denoted by the purple diamond appears for $10^{\circ}\lsim\theta\lsim 100^{\circ}$ and $0.8\lsim J_2/J_1\lsim 100$ in the phase diagram Fig.~\ref{fig:PD}. 
The magnetic space group is $R\bar{3}'$ No. 148.3.1249 in the OG notation~\cite{OG} (No.148.19 $R\bar{3}'$ in the BNS notation~\cite{BNS}). 
The magnetic structure for $\theta=90^{\circ}$ is shown in Fig.~\ref{fig:prp_dia}. 
From $Im\bar{3}$ to the rhombohedral $R\bar{3}'$, one 3-fold axis is singled out as a unique axis. The way of choosing the unique axis among four diagonal directions of cube is four. These four degrees with time-reversal one result in eight, i.e., $4\times 2=8$, which gives eight domains. Our numerically-exact calculation has confirmed this by showing that the degeneracy of the ground state is 8 for $J_1\le 0$.
The total magnetic moment ${\bm J}_{\rm IC}$ and total chirality ${\bm \chi}^{\rm T}$ per icosahedron are zero, ${\bm J}_{\rm IC}={\bf 0}$ and ${\bm \chi}^{\rm T}={\bf 0}$. 
As seen from the unique 3-fold axis (111) in Fig.~\ref{fig:prp_dia}, each magnetic moments on icosahedron swirls and hence we refer to this state as the vortex state. 
The topological charge of the vortex state shown in Fig.~\ref{fig:prp_dia} is $n=0$. It is noted that for $\theta<\theta_{\rm c1}=79.475\dots^{\circ}$ denoted by the dotted line in Fig.~\ref{fig:PD}, the vortex state is characterized by the topological charge $|n|=1$ (for detail see Sect.~\ref{sec:TC_IC}).  
These characteristics are summarized in Table~\ref{tb:mag_state}. 
%Note that the time-reversal state characterized by the opposite sign of $n$, i.e. $n=+1$, is energetically degenerate. 
Note that at the 7th column of Table~\ref{tb:mag_state}, $n$ for the magnetic state shown in each figure listed at the 1st column is denoted.

%%%%% Fig.10 %%%%%%%%%%%%%%%%%%%%%%%%%%%%%%%%%%%%%%%%%%%%%
%%%%%%%%%%%%%%%%%%%%%%%%%%%%%%%%%%%%%%%%%%%%%%%%%%%%%%%%

\begin{figure}[tb]
\includegraphics[width=8cm]{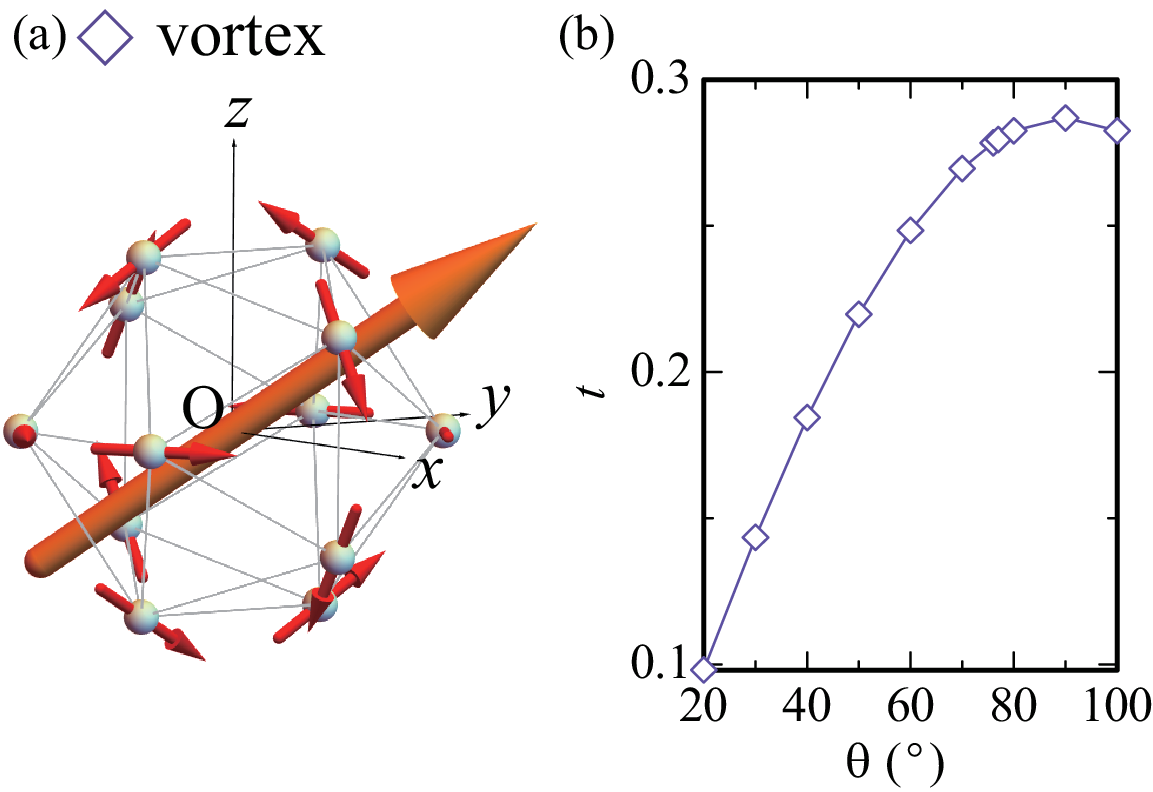}
\caption{
(a) Magnetic toroidal moment vector (orange) in the vortex state with $\theta=90^{\circ}$. 
(b) Anisotropy angle dependence of $t$ in the magnetic toroidal moment ${\bm T}_{\rm IC}=a(t,t,t)$ for the vortex state.  
}
\label{fig:MT_t_th}
\end{figure}

Since the magnetic moments swirl, the vortex state can be characterized by the magnetic toroidal moment on the icosahedron as~\cite{WFK2026}
\begin{eqnarray}
{\bm T}_{\rm IC}=\sum_{i=1}^{12}{\bm r}_{i}\times{\bm J}_i,
\label{eq:MT_IC}
\end{eqnarray}
where ${\bm r}_i$ is the position vector of each rare-earth site in the coordinate where the origin is set to be the center of the icosahedron and ${\bm J}_i$ is the magnetic-moment vector on the $i$th site in the ground state.  
The result of the calculation of Eq.~(\ref{eq:MT_IC}) for the vortex state shown in Fig.~\ref{fig:prp_dia} gives ${\bm T}_{\rm IC}=a(t,t,t)\propto (111)$ [see Fig.~\ref{fig:MT_t_th}(a)]. 
Namely, the cluster magnetic toroidal moment vector is directed along the principal axis of $R\bar{3}'$. 
We have also calculated Eq.~(\ref{eq:MT_IC}) for the ground states denoted by the purple diamonds for $20^{\circ}\le\theta\le 100^{\circ}$ in Fig.~\ref{fig:PD}. The results are shown in Fig.~\ref{fig:MT_t_th}(b). 
As $\theta$ increases, the magnitude of the magnetic toroidal moment increases and has a maximum at $\theta=90^{\circ}$. 
Since the magnetic toroidal moment on each icosahedron is aligned along the principal axis of $R\bar{3}'$ in the vortex order, electromagnetic effect, where the magnetization is induced by applied electric field in the perpendicular direction ${\bm E}\perp{\bm M}$, is expected to appear~\cite{Hayami2014,WFK2026}.  For detail of the theoretical analysis on the 1/1 AC, readers refer to Ref.~\cite{WFK2026}. 

%%%%% Fig.11 %%%%%%%%%%%%%%%%%%%%%%%%%%%%%%%%%%%%%%%%%%%%%
%%%%%%%%%%%%%%%%%%%%%%%%%%%%%%%%%%%%%%%%%%%%%%%%%%%%%%%%

\begin{figure}[tb]
\includegraphics[width=12cm]{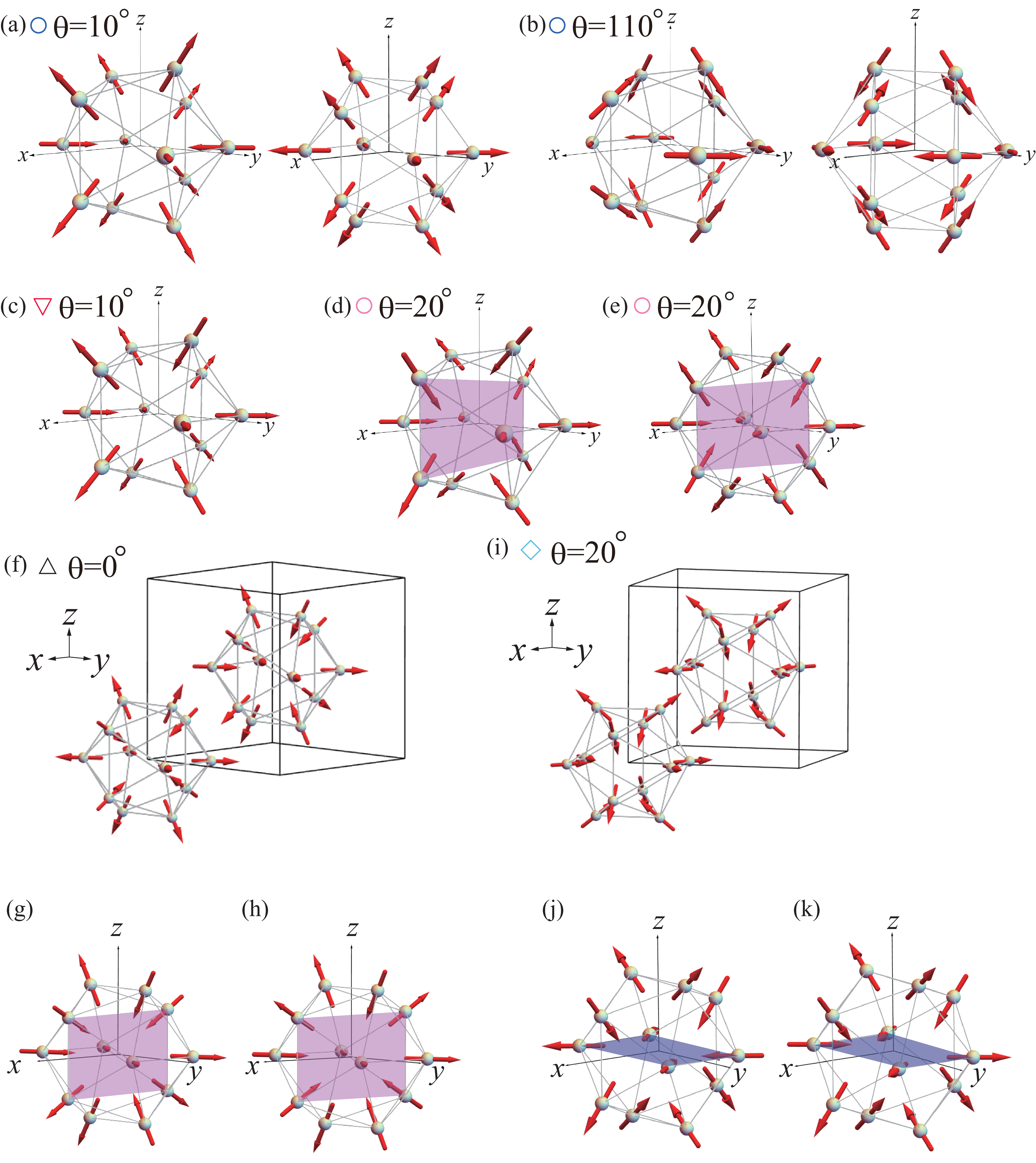}
\caption{
Non-collinear state with anisotropy angles  
(a) $\theta=10^{\circ}$ and (b) $\theta=110^{\circ}$. 
(c) Non-collinear state with anisotropy angle $\theta=10^{\circ}$.
Energetically degenerate states (d) and (e) at the anisotropy angle $\theta=20^{\circ}$ (see text). 
(f) Magnetic state with anisotropy angle $\theta=0^{\circ}$ denoted by the black triangle in Fig.~\ref{fig:PD}.
(g) Magnetic state located at corner icosahedron in the frame box in (f).
(h) Magnetic state located at center icosahedron in the frame box in (f).
(i) Magnetic state with anisotropy angle $\theta=20^{\circ}$ denoted by the light-blue diamond in Fig.~\ref{fig:PD}.
(j) Magnetic state located at center icosahedron in the frame box in (i).
(k) Magnetic state located at corner icosahedron in the frame box in (i).
}
%\label{fig:bl_circ}
\label{fig:FM}
\end{figure}

The magnetic state denoted by the blue circle is stabilized for $\theta\le 10^{\circ}$ and $110^{\circ}\le\theta\le 180^{\circ}$ in the phase diagram Fig.~\ref{fig:PD}. 
The magnetic space group is 
%$I222$ No.23.1.145 
$Im'm'm'$ No. 71.5.625  
in the OG notation~\cite{OG} 
(%No. 23.49 $I222$
$Im'm'm'$ No. 71.537  
in the BNS notation~\cite{BNS}). 
From $Im\bar{3}$ to the orthorhombic 
%$I222$, 
$Im'm'm'$ 
one 2-fold axis is singled out as a unique axis. The way of choosing the unique axis among three orthogonal axes of cube is three. These three degrees with time-reversal one result in six, i.e., $3\times 2=6$, which gives six domains. Our numerically-exact calculation has confirmed this by showing that the degeneracy of the ground state is 6 for $J_1\le 0$.
The magnetic structures for $\theta=10^{\circ}$ and $\theta=110^{\circ}$ are illustrated in 
%Fig.~\ref{fig:bl_circ}(a) and \ref{fig:bl_circ}(b), 
Figs.~\ref{fig:FM}(a) and \ref{fig:FM}(b), 
respectively.
The total magnetic moment on the icosahedron is zero ${\bm J}_{\rm IC}={\bf 0}$ and the total chirality is zero ${\bm \chi}^{\rm T}={\bf 0}$. 
This magnetic state is characterized by the topological charge $n=-1$. 
%This state is obtained by inverting the magnetic moments located at the four vertices of the purple rectangle in Fig.~\ref{fig:bl_circ}(a) from those of the hedgehog state where all the magnetic moments are directed outward from the icosahedron. The energetically degenerate state can also be obtained by inverting the magnetic moments located at the four vertices of the yellow rectangle or the pink rectangle from those of the hedgehog state. Hence, the degeneracy is 3. 
These characteristics are summarized in Table~\ref{tb:mag_state}. 

It is noted that the uniform hedgehog state, where all the magnetic moments are directed outward from the icosahedron (see the right panel of Fig.~\ref{fig:FM}(a))~\cite{WSR}, is energetically degenerate with the magnetic state denoted by the blue circle in Fig.~\ref{fig:PD}. The magnetic space group of the uniform hedgehog state is $Im'\bar{3}'$. The magnetic moment on the icosahedron is zero ${\bm J}_{\rm IC}={\bf 0}$ and the total chirality ${\bf \chi}^{\rm T}={\bf 0}$. The topological charge is $n=+1$. The anti-hedgehog state, where all the magnetic moments are directed inward from the icosahedron~\cite{WSR}, is also energetically degenerate with the hedgehog state. The topological charge of the anti-hedgehog state is $n=-1$. We also plot the energetically degenerate state for $\theta=110^{\circ}$ in the right panel of Fig.~\ref{fig:FM}(b), whose magnetic space group is $Im'\bar{3}'$.

%%%%% Fig.11 %%%%%%%%%%%%%%%%%%%%%%%%%%%%%%%%%%%%%%%%%%%%%
%%%%%%%%%%%%%%%%%%%%%%%%%%%%%%%%%%%%%%%%%%%%%%%%%%%%%%%%%

%The magnetic state denoted by the purple diamond is realized for $90^{\circ}\le\theta\le100^{\circ}$ in the phase diagram of Fig.~\ref{fig:PD}. The magnetic structure for $\theta=20^{\circ}$ and $\theta=90^{\circ}$ are illustrated in Figs.~\ref{fig:pur_dia}(a) and \ref{fig:pur_dia}(b), respectively. The total magnetic moment on the icosahedron is zero ${\bm J}_{\rm IC}={\bf 0}$ and the total chirality is zero ${\bm \chi}^{\rm T}={\bf 0}$. This magnetic state is characterized by the topological charge $n=-1$. The degeneracy of this state is 8. These characteristics are summarized in Table~\ref{tb:mag_state}. 

%%%%% Fig.8 %%%%%%%%%%%%%%%%%%%%%%%%%%%%%%%%%%%%%%%%%%%%%%
%\begin{figure}[b]
%%\includegraphics[width=6cm]{pur_dia_IC_th_90.eps}
%\includegraphics[width=8cm]{pur_dia_IC_th_90.eps}
%\caption{
%Non-collinear state with anisotropy angles 
%(a) $\theta=20^{\circ}$ and 
%(b) $\theta=90^{\circ}$.
%}
%\label{fig:pur_dia}
%\end{figure}
%%%%%%%%%%%%%%%%%%%%%%%%%%%%%%%%%%%%%%%%%%%%%%%%%%%%%%%%%

%-------------------------------------------------------------
\subsubsection{Ferrimagnetic orders}

The magnetic state denoted by the red-inverted triangle appears at $\theta=10^{\circ}$ for small $J_2/J_1$ in the phase diagram Fig.~\ref{fig:PD}.
The magnetic space group is $Im'm'2$ No. 44.4.327 in the OG notation~\cite{OG} ($Im'm'2$ No. 44.232 in the BNS notation~\cite{BNS}). 
The magnetic structure is shown in 
%Fig.~\ref{fig:red_inv_tri}.
Fig.~\ref{fig:FM}(c). 
From space group $Im\bar{3}$ to monoclinic $Im'm'2$, one 2-fold axis is singled out, i.e., 
for the magnetic state shown in 
%Fig.~\ref{fig:red_inv_tri} 
Fig.~\ref{fig:FM}(c) 
the $y$ axis and it turns out the total magnetic moment per icosahedron exists along the unique axis. 
Namely, the total magnetic moment on the icosahedron is ${\bm J}_{\rm IC}=(0.00,0.33,0.00)$ and the total chirality is  ${\bm \chi}^{\rm T}=(0.00,-1.69,0.00)$.  
From $Im\bar{3}$ to orthorhombic $Im'm'2$, the way of choosing the unique axis among three 2-fold axis of cube is three. Then, the way of choosing the second primary axis between the extra two 2-fold-axes is two, which results in six degrees.   
These six degrees with the time-reversal degree result in twelve, i.e., $6\times 2=12$, which gives twelve domains. Our numerically-exact calculation has confirmed this by showing that the degeneracy of the ground state is 12.
These characteristics are summarized in Table~\ref{tb:mag_state}. 

%%%%% Fig.12 %%%%%%%%%%%%%%%%%%%%%%%%%%%%%%%%%%%%%%%%%%%%%
%\begin{figure}[tb]
%\includegraphics[width=6cm]{red_inv_tri_IC_th_10.eps}
%\caption{Non-collinear state with anisotropy angle $\theta=10^{\circ}$.}
%\label{fig:red_inv_tri}
%\end{figure}
%%%%%%%%%%%%%%%%%%%%%%%%%%%%%%%%%%%%%%%%%%%%%%%%%%%%%%%%%

The magnetic state denoted by the pink circle appears at $\theta=20^{\circ}$ for small $J_2/J_1$ in the phase diagram Fig.~\ref{fig:PD}.
%The magnetic structure is illustrated in Fig.~\ref{fig:pink_circ}(a). 
The magnetic space group is $I2'2'2$ No. 23.3.147 in the OG notation~\cite{OG} ($I2'2'2$ No. 23.51 in the BNS notation~\cite{BNS}). 
For the magnetic state shown in 
%Fig.~\ref{fig:pink_circ}(a), 
Fig.~\ref{fig:FM}(d) 
the magnetic structure is invariant under the $180^{\circ}$ rotation around the $y$ axis while the magnetic structure is invariant under the $180^{\circ}$ rotation around the $x$ axis or the $y$ axis with inverting all the magnetic moments. It turns out that the total magnetic moment per icosahedron and the total chirality vectors are directed to the unique axis, i.e., the 2-fold axis direction without the magnetic-moment inversion. Namely, it is the $y$ axis for 
%Fig.~\ref{fig:pink_circ}(a) 
Fig.~\ref{fig:FM}(d) 
and  
the total magnetic moment on the icosahedron is ${\bm J}_{\rm IC}=(0.00,-0.66,0.00)$ and the total chirality is  ${\bm \chi}^{\rm T}=(0.00,-3.12,0.00)$. 
From $Im\bar{3}$ to orthorhombic $I2'2'2$, the way of choosing the unique axis among three 2-fold axes of cube is three. This three degrees with the time-reversal degree resuts in six, i.e., $3\times 2=6$, which gives six domains. 

We confirmed that the magnetic structure illustrated in 
%Fig.~\ref{fig:pink_circ}(b) 
Fig.~\ref{fig:FM}(d)  
is energetically degenerate with the state shown in 
%Fig.~\ref{fig:pink_circ}(a).
Fig.~\ref{fig:FM}(d). 
This state is obtained by inverting all the magnetic moments located at the four vertices of the pink rectangle in 
%Fig.~\ref{fig:pink_circ}(a).
Fig.~\ref{fig:FM}(d) 
This state has the same ${\bm J}_{\rm IC}$ and ${\bm \chi}^{\rm T}$ as those of the state shown  in 
%Fig.~\ref{fig:pink_circ}(a). 
Fig.~\ref{fig:FM}(d). 
%The other states are obtained by permutating each $(x,y,z)$ axis as $(y,z,x)$ axis and $(z,x,y)$ axis in Figs.~\ref{fig:pink_circ}(a) and \ref{fig:pink_circ}(b). 
%Hence, the $3\times 2=6$ states are energetically degenerate. Since the states obtained by inverting all the magnetic moments are also energetically degenerate, the total degeneracy is 12.  
%We have obtained twelve degeneracy for the ground state by our numerically-exact calculation.

Our numerically-exact calculation has shown that the degeneracy of the ground state denoted by the pink circle is 12, which confirms the degeneracy 6 in 
%Fig.~\ref{fig:pink_circ}(a) 
Fig.~\ref{fig:FM}(d)  
and 6 in 
%Fig.~\ref{fig:pink_circ}(b). 
Fig.~\ref{fig:FM}(e).  
These characteristics are summarized in Table~\ref{tb:mag_state}.

%%%%% Fig.13 %%%%%%%%%%%%%%%%%%%%%%%%%%%%%%%%%%%%%%%%%%%%%%
%\begin{figure}[bt]
%%\includegraphics[width=6cm]{pink_circ_IC_th_20.eps}
%\includegraphics[width=8cm]{pink_circ_IC_th_20.eps}
%\caption{
%Energetically degenerate states (a) and (b) at the anisotropy angle $\theta=20^{\circ}$ (see text). 
%}
%\label{fig:pink_circ}
%\end{figure}
%%%%%%%%%%%%%%%%%%%%%%%%%%%%%%%%%%%%%%%%%%%%%%%%%%%%%%%%%

%\onecolumngrid

%%%% Table 1 %%%%%%%%%%%%%%%%%%%%%%%%%%%%%%%%%%%%
%\begin{table}
\begin{table}
\caption{
Characteristics of the magnetically ordered states in the 1/1 AC. 
The symbol in the 1st column denotes the magnetic state shown in the ground-state phase diagram Fig.~\ref{fig:PD}. 
%As for the distribution, u indicates the uniform distribution and s indicates the staggered distribution in the bcc unit cell of the 1/1 AC.
The second column denotes the magnetic propagation vector ${\bm q}=\frac{2\pi}{a}(hkl)$ with $a$ being the lattice constant. 
The third column denotes the magnetic space group. 
The fourth column denotes the degeneracy of the ground state. 
On the fifth, sixth and seventh column, we list ${\bm J}_{\rm IC}$, ${\bm \chi}^{\rm T}$ and $n$ for the center icosahedron in the frame box of the unit cell and $-{\bm J}_{\rm IC}$, $-{\bm \chi}^{\rm T}$, and $-n$ are realized for the corner icosahedron in the antiferromagnetic orders. For degenerate ground states, ${\bm J}_{\rm IC}$, ${\bm \chi}^{\rm T}$, $n$ are not listed for each time-reversal state. 
%The eighth column denotes the topological charge $n$. 
%The IC1 (IC2) is the icosahedron located at the corner (center) of the bcc unit cell.
}
\label{tb:mag_state}
%\begin{ruledtabular}
\renewcommand{\arraystretch}{1.2}
\resizebox{\textwidth}{!}{%
%\begin{tabular}{lccccccdr}
\begin{tabular}{lcccccccr}
\hline
\textrm{state}&  $(hkl)$ &mag.  & 
\textrm{degn.} & \textrm{${\bm J}_{\rm IC}$} & \textrm{${\bm \chi}^{\rm T}$} &
%\multicolumn{2}{c}{$n$}
$n$
\\
& & sp. gr.  & & & &
%\multicolumn{1}{c}{\textrm{IC1}}
%IC1
 
%&
%\textrm{IC2}
\\
\hline
%\colrule
%{\color{red}${\triangledown}$} 
red inverted triangle 
%Fig.~\ref{fig:red_inv_tri}
Fig.~\ref{fig:FM}(c) 
& (000) & $Im'm'2$  & 12 & $(0.00, 0.33, 0.00)^{\rm a}$
%\footnote{$\theta=10^{\circ}$ case.} 
& $(0.00,-1.69,0.00)$ & -2  \\
&  & & & $(-0.33,0.00,0.00)$ & $(1.69,0.00,0.00)$ & -2 \\
&  & & & $(0.33,0.00,0.00)$ & $(-1.69,0.00,0.00)$ & -2 \\
&  & & & $(0.00,0.00,0.33)$ & $(0.00,0.00,-1.69)$ & -2 \\
&  & & & $(0.00,-0.33,0.00)$ & $(0.00,1.69,0.00)$ & -2  \\
&  & & & $(0.00,0.00,-0.33)$ & $(0.00,0.00,1.69)$ & -2 \\
%& & & & (0.00,0.33,0.00) & ${\bf 0}$ & -2 & -2 \\
\hline
%{\color{blue}${\bigcirc}$} 
blue circle
%Fig.~\ref{fig:bl_circ} 
Fig.~\ref{fig:FM}(a) 
& (000) & 
%$I222$ 
$Im'm'm'$ 
& 6 & ${\bf 0}$ & ${\bf 0}$ & -1$^{\rm b}$
%\footnote{$\theta=10^{\circ}$ case.} 
\\
%\hline
%{\color{brown}${\triangle}$} Fig.~\ref{fig:br_tri}& (000) & $I222$   & 2 & ${\bf 0}$ & ${\bf 0}$ & -1 \\
\hline
%{\color{Magenta}${\bigcirc}$} 
pink circle 
%Fig.~\ref{fig:pink_circ}
Fig.~\ref{fig:FM}(d) 
& (000) & $I2'2'2$  & 12 & $(0.00,-0.66, 0.00)$ & $(0.00,-3.12, 0.00)^{\rm c}$
%$\footnote{$\theta=20^{\circ}$ case (see text and Fig.~\ref{fig:pink_circ}).} 
& 0 \\
% & & & $(0.00,-0.66, 0.00)$ & $(0.00,-4.82,0.00)$ & 0 & 0 \\
&  & & & $(0.00, 0.00,-0.66)$ & $(0.00, 0.00,-3.12)$ & 0 \\
% & & & $(0.00, 0.00,-0.66)$ & $(0.00, 0.00,-4.82)$ & 0 & 0 \\
& & & & $(-0.66, 0.00, 0.00)$ & $(-3.12, 0.00, 0.00)$ & 0 \\
% & & & $(-0.66, 0.00, 0.00)$ & $(-4.82, 0.00, 0.00)$ & 0 & 0 \\
\hline
%\scalebox{1.8}{\color{Violet}${\diamond}$} 
purple diamond 
Fig.~\ref{fig:prp_dia}& (000) & $R\bar{3}'$  & 8 & ${\bm 0}$ & ${\bm 0}$  & 0$^{\rm d}$
%\footnote{$\theta=90^{\circ}$ case. For detail see Sect.~\ref{sec:TC_IC}.} 
\\
\hline
%{\color{green}$\blacktriangledown$} 
filled green inverted triangle 
%Fig.~\ref{fig:grn_inv_tri}(a) 
Fig.~\ref{fig:AF}(a) 
& (111) & $R_{\rm R}\bar{3}$  & 8 & $(-0.58,-0.58,-0.58)$ & $(-0.06,-0.06,-0.06)^{\rm e}$
%\footnote{$\theta=50^{\circ}$ case.} 
& 0 \\
& & & & $(0.58,-0.58,-0.58)$ & $(0.06,-0.06,-0.06)$ &0 \\
& & & & $(-0.58,-0.58,0.58)$ & $(-0.06,-0.06,0.06)$ &0 \\
& & & & $(0.58,-0.58,0.58)$ & $(0.06,-0.06,0.06)$ & 0 \\
\hline
%{\color{orange}$\blacksquare$ }
filled orange square 
%Fig.~\ref{fig:or_sq}(b)
Fig.~\ref{fig:AF}(b) 
& (111) & $I_{\rm P}m'\bar{3}'$  & 2 & {\bf 0} & {\bf 0} & 3$^{\rm f}$
%\footnote{$\theta=90^{\circ}$ case. For detail, see Sect.~\ref{sec:TC_IC}.
%$79.48^{\circ}<\theta<103.57^{\circ}$. $n=+1$ for $\theta<79.48^{\circ}$. $n=+3$ for $103.57^{\circ}<\theta<127.67^{\circ}$. $n=-1$ for $\theta>127.67^{\circ}$ (see text).
%} 
\\
\hline
%{\color{cyan}œ} 
filled light-blue circle 
%Fig.~\ref{fig:lbl_circ}(c)
Fig.~\ref{fig:AF}(c)  
& (111) & $R_{\rm R}\bar{3}$   & 8 & $(0.12, 0.12, 0.12)$ & $(-0.22,-0.22,-0.22)^{\rm g}$
%\footnote{$\theta=150^{\circ}$ case.} 
& 0 \\
& & & & $(0.12, 0.12,-0.12)$ & $(-0.22,-0.22,0.22)$ & 0 \\
& & & & $(0.12,-0.12, 0.12)$ & $(-0.22,0.22,-0.22)$ & 0 \\ 
& & & & $(-0.12, 0.12, 0.12)$ & $(0.22,-0.22,-0.22)$ & 0 \\
\hline
\end{tabular}
}
$^{\rm a}$\footnotesize{$\theta=10^{\circ}$ case.} 
$^{\rm b}$\footnotesize{$\theta=10^{\circ}$ case.} 
$^{\rm c}$\footnotesize{$\theta=20^{\circ}$ case [see text and 
%Fig.~\ref{fig:pink_circ}).
Fig.~\ref{fig:FM}(d)]
} 
$^{\rm d}$\footnotesize{$\theta=90^{\circ}$ case. For detail see Sect.~\ref{sec:TC_IC}.}
$^{\rm e}$\footnotesize{$\theta=50^{\circ}$ case.} 
$^{\rm f}$\footnotesize{$\theta=90^{\circ}$ case. For detail, see Sect.~\ref{sec:TC_IC}.}
$^{\rm g}$\footnotesize{$\theta=150^{\circ}$ case.} 
\renewcommand{\arraystretch}{1}
%\end{ruledtabular}
\end{table}
%\end{table}
%%%%%%%%%%%%%%%%%%%%%%%%%%%%%%%%%%%%%%%%%%%%%%%%%

%\twocolumngrid

%-------------------------------------------------------

The magnetic state denoted by the black triangle appears for $0^{\circ}\lsim\theta\lsim 10^{\circ}$ and small $J_2/J_1$ in the phase diagram Fig.~\ref{fig:PD}. 
The magnetic space group is $P2'/m'$ No. 10.5.53 in the OG notation ($P2'/m'$ No. 10.46 in the BNS notation). 
%From space group $Im\bar{3}$ to monoclinic  $Cm'$, one 2-fold axis is singled out as the unique axis and it turns out that the net magnetic moment per icosahedron vanishes along the unique axis. 
The magnetic structure with anisotropy angle $\theta=0^{\circ}$ is shown in 
%Fig.~\ref{fig:gr_tri}(a). 
Fig.~\ref{fig:FM}(f). 
The magnetic structure is invariant under the mirror operation with respect to the mirror plane perpendicular to the $x$ axis and the time reversal (magnetic-moment inversion) operation. 
The magnetic structure is also invariant under the 2-fold axis rotation around the $x$ axis with the time reversal operation. 
%Figures~\ref{fig:gr_tri}(b) and \ref{fig:gr_tri}(c) 
Figures~\ref{fig:FM}(g) and \ref{fig:FM}(h) 
show the magnetic states located at the center icosahedron and corner icosahedron in the frame box shown in 
%Fig.~\ref{fig:gr_tri}(a), 
Fig.~\ref{fig:FM}(f), 
respectively. 
By inverting the magnetic moments located at four vertices of the magenta rectangular in 
%Fig.~\ref{fig:gr_tri}(b), 
Fig.~\ref{fig:FM}(g), 
the magnetic state shown in 
%Fig.~\ref{fig:gr_tri}(c) 
Fig.~\ref{fig:FM}(h) 
is obtained. 
The total magnetic moment per icosahedron is calculated as ${\bm J}_{\rm IC}=(0.00,1.30,-2.10)$ for 
%Fig.~\ref{fig:gr_tri}(b) 
Fig.~\ref{fig:FM}(g) 
and ${\bm J}_{\rm IC}=(0.00,1.30,2.10)$ for 
%Fig.~\ref{fig:gr_tri}(c), 
Fig.~\ref{fig:FM}(h), 
which result in the net moment along the $y$ direction. 
The total chirality are obtained as ${\bm \chi}^{\rm T}=(0.00,-3.94,6.35)$ and  ${\bm \chi}^{\rm T}=(0.00,-3.94,-6.35)$ for 
%Figs.~\ref{fig:gr_tri}(b) and \ref{fig:gr_tri}(c), 
Figs.~\ref{fig:FM}(g) and \ref{fig:FM}(h)   
respectively, which result in the net total chirality along the $-y$ direction. Hence, topological Hall effect is expected to appear in the electron conductivity $\sigma_{zx}$ on the basis of Eq.~(\ref{eq:sigma}). 
The topological charge for both states shown in 
%Figs.~\ref{fig:gr_tri}(b) and \ref{fig:gr_tri}(c) 
Figs.~\ref{fig:FM}(g) and \ref{fig:FM}(h) 
is zero $n=0$. 
The degeneracy of the ground state is 12 for $J_2<0$. 
These characteristics are summarized in Table~\ref{tb:mag_state_2}. Note that time-reversal state where all the magnetic moments are inverted is energetically degenerate so that each ground state with $-{\bm J}_{\rm IC}$ and $-{\bm \chi}^{\rm T}$ exists as the degenerated state. We note that the degeneracy for $J_2=0$ is 192, where the magnetic state shown in 
%Fig.~\ref{fig:gr_tri}(a) 
Fig.~\ref{fig:FM}(f)  
is included in the degenerated states.

%%%%% Fig.14 %%%%%%%%%%%%%%%%%%%%%%%%%%%%%%%%%%%%%%%%%%%%%%
%\begin{figure}[tb]
%\includegraphics[width=8cm]{gr_tri_IC_th_0.eps}
%\caption{
%(a) Magnetic state with anisotropy angle $\theta=0^{\circ}$ denoted by the black triangle in Fig.~\ref{fig:PD}.
%(b) Magnetic state located at corner icosahedron in the frame box in (a).
%(c) Magnetic state located at center icosahedron in the frame box in (a).
%}
%\label{fig:gr_tri}
%\end{figure}
%%%%%%%%%%%%%%%%%%%%%%%%%%%%%%%%%%%%%%%%%%%%%%%%%%%%%%%%%

The magnetic state denoted by the light-blue diamond appears for $20^{\circ}\lsim\theta\lsim 30^{\circ}$ and small $J_2/J_1$ in the phase diagram Fig.~\ref{fig:PD}. 
The magnetic space group is $P\bar{1}$ No. 2.1.4 in the OG notation ($P\bar{1}$ No. 2.4 in the BNS notation). 
%From space group $Im\bar{3}$ to monoclinic  $Cm$, one 2-fold axis is singled out and it turns out that the net magnetic moment per icosahedron exists along the unique axis. 
The magnetic structure with anisotropy angle $\theta=20^{\circ}$ is shown in 
%Fig.~\ref{fig:gr_dia}(a). 
Fig.~\ref{fig:FM}(i) 
The magnetic structure is invariant under spatial inversion with the magnetic moment being unchanged.  
The magnetic states located at the center icosahedron and corner icosahedron in the frame box of 
%Fig.~\ref{fig:gr_dia}(a) 
Fig.~\ref{fig:FM}(i)  
are shown in 
%Figs.~\ref{fig:gr_dia}(b) and \ref{fig:gr_dia}(c), 
Figs.~\ref{fig:FM}(j)  and \ref{fig:FM}(k) 
respectively.
By inverting the magnetic moments located at four vertices of the purple rectangular in 
%Fig.~\ref{fig:gr_dia}(b), 
Fig.~\ref{fig:FM}(j), 
the magnetic state shown in 
%Fig.~\ref{fig:gr_dia}(c) 
Fig.~\ref{fig:FM}(k)  
is obtained. 
The total magnetic moment per icosahedron is calculated as ${\bm J}_{\rm IC}=(0.00,-2.48,-0.66)$ for 
%Fig.~\ref{fig:gr_tri}(b) 
Fig.~\ref{fig:FM}(j) 
and ${\bm J}_{\rm IC}=(0.00,2.48,-0.66)$ for 
%Fig.~\ref{fig:gr_dia}(c), 
Fig.~\ref{fig:FM}(k)  
which result in the net moment along the $-z$ direction.
The total chirality are obtained as ${\bm \chi}^{\rm T}=(0.00,4.21,1.37)$ and  ${\bm \chi}^{\rm T}=(0.00,-4.21,1.37)$ for 
%Figs.~\ref{fig:gr_dia}(b) and \ref{fig:gr_dia}(c), 
Figs.~\ref{fig:FM}(j) and \ref{fig:FM}(k), 
%respectively, which result in the cancellation of the net total chirality.
respectively, which result in the net total chirality along the $z$ direction.
The topological charge for both states shown in 
%Figs~\ref{fig:gr_dia}(b) and \ref{fig:gr_dia}(C) 
Figs.~\ref{fig:FM}(j) and \ref{fig:FM}(k) 
is zero $n=0$. 
The degeneracy of the ground state is 12 for $J_2<0$.
These characteristics are summarized in Table~\ref{tb:mag_state_2}. 
We note that the degeneracy for $J_2=0$ is 192, where the magnetic state shown in 
%Fig.~\ref{fig:gr_dia}(a) 
Fig.~\ref{fig:FM}(i)  
is included in the degenerated states. 

%%%%% Fig.15 %%%%%%%%%%%%%%%%%%%%%%%%%%%%%%%%%%%%%%%%%%%%%%
%\begin{figure}[tb]
%\includegraphics[width=8cm]{gr_dia_IC_th_20.eps}
%\caption{
%(a) Magnetic state with anisotropy angle $\theta=20^{\circ}$ denoted by the light-blue diamond in Fig.~\ref{fig:PD}.
%(b) Magnetic state located at center icosahedron in the frame box in (a).
%(c) Magnetic state located at corner icosahedron in the frame box in (a).
%}
%\label{fig:gr_dia}
%\end{figure}
%%%%%%%%%%%%%%%%%%%%%%%%%%%%%%%%%%%%%%%%%%%%%%%%%%%%%%%%%

%\onecolumngrid

%%%% Table 2 %%%%%%%%%%%%%%%%%%%%%%%%%%%%%%%%%%%%
\begin{table}
\caption{
Characteristics of the ferrimagnetic states with expanded unit cell of the 1/1 AC. The symbol in the 1st column denotes the magnetic state shown in the ground-state phase diagram Fig.~\ref{fig:PD}. 
The second column denotes the magnetic propagation vector ${\bm q}=\frac{2\pi}{a}(hkl)$ with $a$ being the lattice constant. 
The third column denotes the magnetic space group. 
The fourth column denotes the degeneracy of the ground state. 
The fifth and sixth columns denote ${\bm J}_{\rm IC}$ for IC1 and IC2, respectively.
The seventh and eighth columns denote ${\bm \chi}^{\rm T}$ for IC1 and IC2, respectively.
The IC1 (IC2) is the icosahedron located at the center (corner) in the frame box in Fig.~\ref{fig:lattice}. 
}
\label{tb:mag_state_2}
%\begin{ruledtabular}
\renewcommand{\arraystretch}{1.2}
\resizebox{\textwidth}{!}{%
%\begin{tabular}{lccccccdr}
\begin{tabular}{lcccccccr}
\hline
\textrm{state}&  $(hkl)$ &mag.  & 
\textrm{degn.} & \multicolumn{2}{c}{${\bm J}_{\rm IC}$} &  \multicolumn{2}{c}{${\bm \chi}^{\rm T}$}  
\\
& & sp. gr.  & & IC1 & IC2 & IC1  & IC2  
\\
%\colrule
\hline
%----------------
%\hline 
\color{black}${\triangle}$ 
%Fig.~\ref{fig:gr_tri} 
Fig.~\ref{fig:FM}(f)
& (000) & $P2'/m'$ & 12 &  $(0.00,1.30,-2.10)$ & $(0.00,1.30,2.10)$ & $(0.00,-3.94,6.35)$ & $(0.00,-3.94,-6.35)^{\rm a}$
%\footnote{$\theta=0^{\circ}$ case.} 
\\
 & & & & $(0.00,1.30,2.10)$ &  $(0.00,1.30,-2.10)$ & $(0.00,-3.94,-6.35)$   & $(0.00,-3.94,6.35)$
\\
 & & & & $(-2.10,0.00,1.30)$ & $(2.10,0.00,1.30)$  & $(6.35,0.00,-3.94)$  & $(-6.35,0.00,-3.94)$
\\
 & & & & $(1.30,2.10,0.00)$ & $(1.30,-2.10,0.00)$ & $(-3.94,-6.35,0.00)$ & $(-3.94,6.35,0.00)$
\\
 & & & & $(2.10,0.00,1.30)$ & $(-2.10,0.00,1.30)$ & $(-6.35,0.00,-3.94)$ & $(6.35,0.00,-3.94)$
\\
 & & & & $(1.30,-2.10,0.00)$ & $(1.30,2.10,0.00)$ & $(-3.94,6.35,0.00)$ & $(-3.94,-6.35,0.00)$ 
\\
\hline
\scalebox{1.8}{\color{cyan}${\diamond}$} 
%Fig.~\ref{fig:gr_dia} 
Fig.~\ref{fig:FM}(i)  
& (000) & $P\bar{1}$ & 12 &  $(0.00,2.48,0.66)$ & $(0.00,-2.48,0.66)$ & $(0.00,-4.21,1.37)$ & $(0.00,4.21,1.37)^{\rm b}$
%\footnote{$\theta=20^{\circ}$ case.} 
\\
 & & & & $(2.48,0.66,0.00)$ & $(-2.48,0.66,0.00)$ & $(-4.21,1.37,0.00)$ & $(4.21,1.37,0.00)$
\\
 & & & & $(0.66,0.00,2.48)$ & $(0.66,0.00,-2.48)$ & $(1.37,0.00,-4.21)$ & $(1.37,0.00,4.21)$
\\
 & & & & $(0.00,-2.48,0.66)$ & $( 0.00,2.48,0.66)$ & $(0.00,4.21,1.37)$ & $(0.00,-4.21,1.37)$
\\
 & & & & $(0.66,0.00,-2.48)$ & $(0.66,0.00,2.48)$ & $(1.37,0.00,4.21)$ & $(1.37,0.00,-4.21)$
\\
 & & & & $(-2.48,0.66,0.00)$ & $(2.48,0.66,0.00)$ & $(4.21,1.37,0.00)$ & $(-4.21,1.37,0.00)$
\\
\hline
\end{tabular}
}
$^{\rm a}$\footnotesize{$\theta=0^{\circ}$ case.}
$^{\rm b}$\footnotesize{$\theta=20^{\circ}$ case.}
\renewcommand{\arraystretch}{1}
%\end{ruledtabular}
\end{table}
%\end{table}
%%%%%%%%%%%%%%%%%%%%%%%%%%%%%%%%%%%%%%%%%%%%%%%%%

%\twocolumngrid

%------------------------------------------------------------------
%\section{Discussion}

\section{Topological charge on icosahedron}
\label{sec:TC_IC}

We have defined the topological charge on the icosahedron in Eq.~(\ref{eq:tp})~\cite{WPNAS}. In the ground-state phase diagram shown in Fig.~\ref{fig:PD}, the colored regions denote magnetic orders characterized by the topological charge. Here, we analyze the topological charge of the vortex state and anti-whirling state as a function of $\theta$, which results in characteristic anisotropy angles $\theta_{\rm c1}$, $\theta_{\rm c2}$, and $\theta_{\rm c3}$.

\begin{figure}[tb]
\includegraphics[width=8cm]{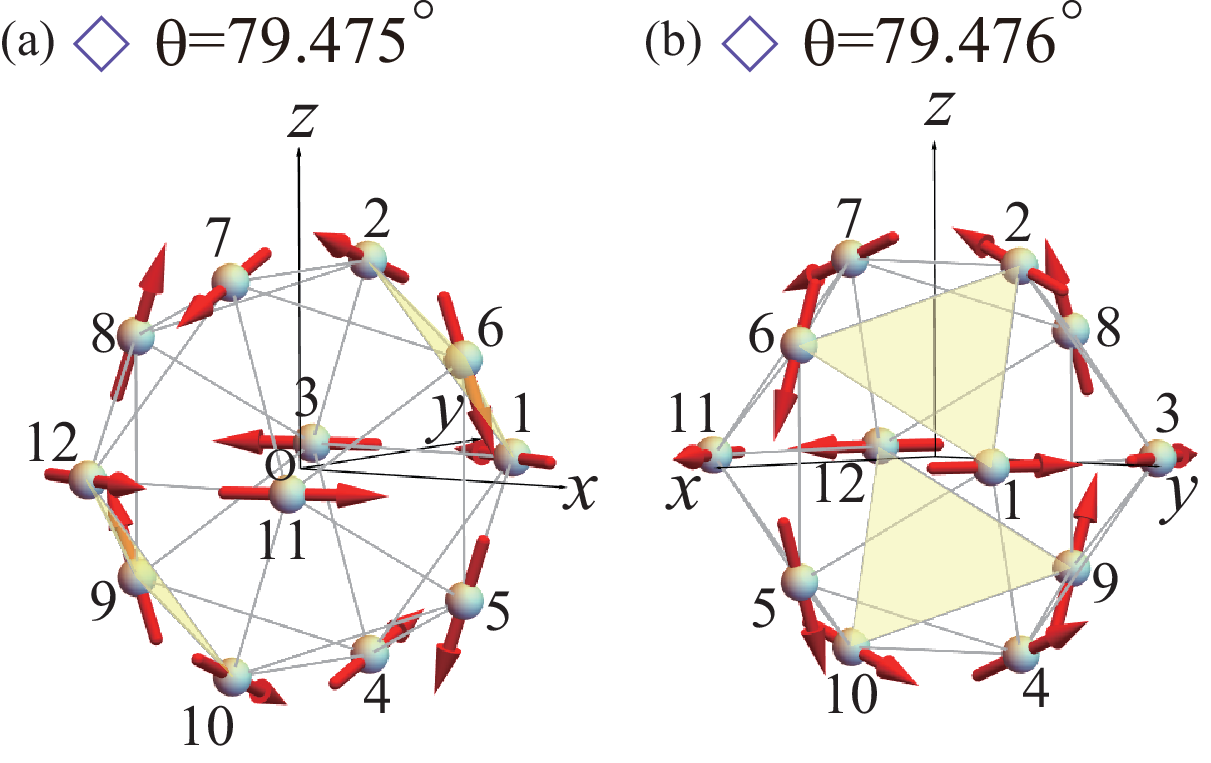}
\caption{
Vortex state at (a) $\theta=79.475^{\circ}=\theta_{\rm c1}^{-}$ and (b) $\theta=79.476^{\circ}=\theta_{\rm c1}^{+}$. The number labels the rare-earth sites located at 12 vertices of icosahedron. 
}
\label{fig:vortex_th_c1}
\end{figure}

For the vortex state shown in Fig.~\ref{fig:prp_dia}, the topological charge $n$ defined by Eq.~(\ref{eq:tp}) is $n=0$ for $\theta>\theta_{\rm c1}$ where $\theta_{\rm c1}$ is identified as $\theta_{\rm c1}=79.475\dots^{\circ}$, while $n=-1$ for $\theta<\theta_{\rm c1}$. 
At $\theta=\theta_{\rm c1}$, in the denominator of the right hand side (r.h.s) of Eq.~(\ref{eq:SI_tri}), the following relation
\begin{eqnarray}
{\bm J}_i\cdot{\bm J}_j+{\bm J}_j\cdot{\bm J}_k+{\bm J}_k\cdot{\bm J}_i=-1
\label{eq:inp_tri}
\end{eqnarray}
holds for the two $ijk$ triangular surfaces on icosahedron which are perpendicular to the principal axis of rhombohedral $R\bar{3}'$. For instance, the principal axis of the vortex state shown in Figs.~\ref{fig:vortex_th_c1}(a) and \ref{fig:vortex_th_c1}(b) is the (111) direction, i.e., each magnetic moment swirls around (111) and the magnetic toroidal moment in Eq.~(\ref{eq:MT_IC}) is directed as ${\bm T}_{\rm IC}\parallel (111)$. Since the denominator of the r.h.s. of Eq.~(\ref{eq:SI_tri}) becomes zero, the solid angle subtended by the three magnetic moments yields $\Omega_{ijk}=\pm\pi$ ($+$ for $\theta_{\rm c1}^{+}\equiv\theta_{\rm c1}+\delta$ with $\delta$ being positive definite and $-$ for $\theta_{\rm c1}^{-}\equiv\theta_{\rm c1}-\delta$). 
Then, the solid angle of the $(i,j,k)=(1,2,6)$ as well as $(9,10,12)$ surface on the icosahedron changes from $\frac{\Omega_{ijk}}{4\pi}=-\frac{1}{4}$ to $+\frac{1}{4}$ as $\theta$ increases from $\theta_{\rm c1}^{-}$ to $\theta_{\rm c1}^{+}$. Hence, the change in the topological charge is $(\frac{1}{4}-(-\frac{1}{4}))\times 2=+1$ so that the topological charge $n=-1$ for $\theta<\theta_{\rm c1}$ changes as  $n=-1+1=0$ for $\theta>\theta_{\rm c1}$.  
 
%%%%% Fig.17 %%%%%%%%%%%%%%%%%%%%%%%%%%%%%%%%%%%%%%%%%%%%%%
%%%%%%%%%%%%%%%%%%%%%%%%%%%%%%%%%%%%%%%%%%%%%%%%%%%%%%%%%

\begin{figure}[tb]
\includegraphics[width=8cm]{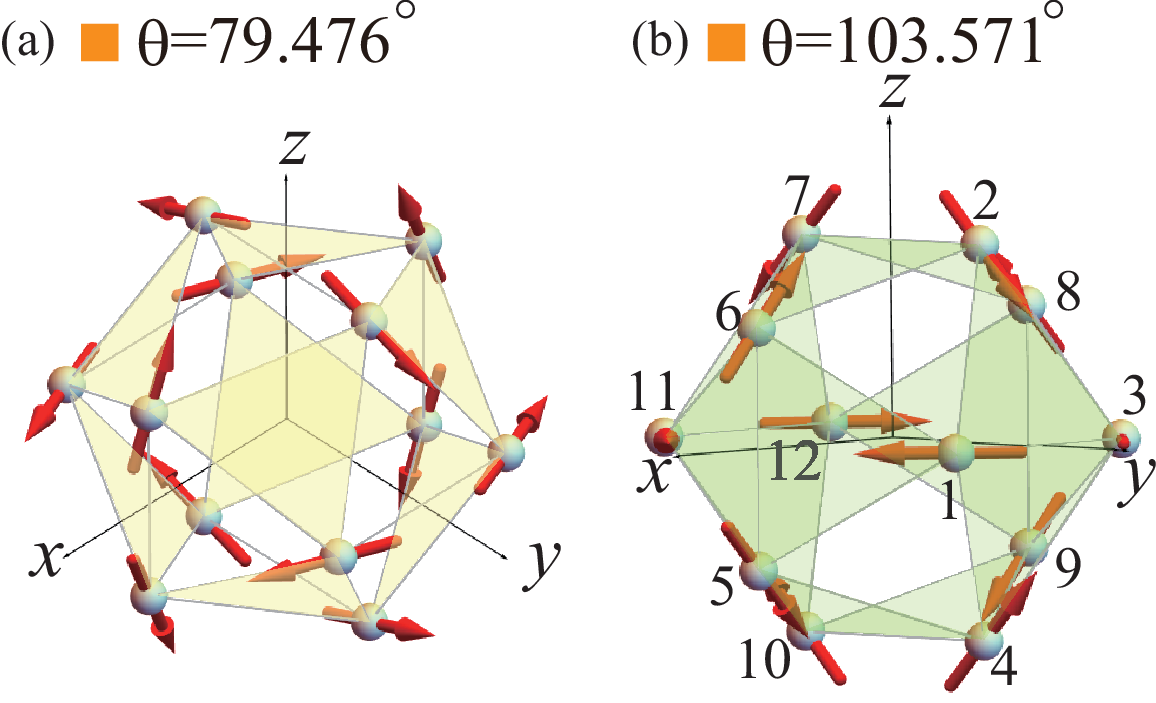}
\caption{
(a) Anti-whirling state at $\theta=\theta_{\rm c1}^{+}$. 
(b) Whirling state at $\theta=103.571^{\circ}=\theta_{\rm c2}^{+}$. The number labels the rare-earth sites located at 12 vertices of icosahedron. 
}
\label{fig:AW_th_c1}
\end{figure}

The anti-whirling state shown in 
%Fig.~\ref{fig:AW_th_c1}(a) 
Fig.~\ref{fig:AF}(b) 
at $\theta=79.476^{\circ}=\theta_{\rm c1}^{+}$ is characterized by the topological charge $n=-3$.  
At $\theta=\theta_{\rm c1}$, Eq.~(\ref{eq:inp_tri}) is satisfied for the eight $ijk$ triangular surfaces on icosahedron colored in light yellow in Fig.~\ref{fig:AW_th_c1}(a) each of which is perpendicular to the 3-fold axis direction of cubic $I_{\rm P}m'\bar{3}'$, i.e., $(\pm1,\pm1,\pm1)$. Then, the solid angle of the 8 surfaces on icosahedron changes from $\frac{\Omega_{ijk}}{4\pi}=-\frac{1}{4}$ to $\frac{1}{4}$ as $\theta$ decreases from $\theta_{\rm c1}^{+}$ to $\theta_{\rm c1}^{-}$. Hence, the change in the topological charge is $(\frac{1}{4}-(-\frac{1}{4}))\times 8=+4$ so that the topological charge $n=-3$ for $\theta>\theta_{\rm c1}$ changes as  $n=-3+4=+1$ for $\theta<\theta_{\rm c1}$.  

As for the whirling state where all the magnetic moments are inverted in Fig.~\ref{fig:AW_th_c1}(a), the topological charge $n=+3$ for $\theta>\theta_{{\rm c}1}$ changes as $n=-1$ for $\theta<\theta_{{\rm c}1}$.

%%%%% Fig.16 %%%%%%%%%%%%%%%%%%%%%%%%%%%%%%%%%%%%%%%%%%%%%%
%%%%%%%%%%%%%%%%%%%%%%%%%%%%%%%%%%%%%%%%%%%%%%%%%%%%%%%%%

We have found that at $\theta=\theta_{\rm c2}=103.570\dots^{\circ}$, coplanar alignments of the magnetic moments appear at the $ijk$ triangular surfaces perpendicular to the eight diagonal directions of cube, i.e., $(\pm1,\pm1,\pm1)$ [see the whirling state for $\theta=103.571^{\circ}=\theta_{\rm c2}^{+}$ shown in Fig.~\ref{fig:AW_th_c1}(b)]. Hence, the scalar chirality vanishes as $\chi_{ijk}=0$ in Eq.~(\ref{eq:SC}) so that the solid angle becomes zero as $\Omega_{ijk}=0$ in Eq.~(\ref{eq:SI_tri}) for these 8 triangular surfaces. For extra 12 triangular surfaces on icosahedron colored in light green surfaces in Fig.~\ref{fig:AW_th_c1}(b), we have confirmed that Eq.~(\ref{eq:inp_tri}) is satisfied at $\theta=\theta_{\rm c2}$.  Then, the solid angle changes from $\frac{\Omega_{ijk}}{4\pi}=-\frac{1}{4}$ to $+\frac{1}{4}$ as $\theta$ increases from $\theta_{\rm c2}^{-}$ to $\theta_{\rm c2}^{+}$. Hence, the change in the topological charge is $(\frac{1}{4}-(-\frac{1}{4}))\times 12=+6$ so that the topological charge $n=-3$ for $\theta_{\rm c1}<\theta<\theta_{\rm c2}$ changes as $n=-3+6=+3$ for $\theta>\theta_{\rm c2}$.  

As for the anti-whirling state where all the magnetic moments are inverted in Fig.~\ref{fig:AW_th_c1}(b), the topological charge $n=-3$ for $\theta>\theta_{{\rm c}2}$ changes as $n=+3$ for $\theta_{{\rm c}1}<\theta<\theta_{{\rm c}2}$. 

We have also found that at $\theta=\theta_{\rm c3}=127.665\dots^{\circ}$, Eq.~(\ref{eq:inp_tri}) is satisfied for the eight $ijk$ triangular surfaces on icosahedron each of which is perpendicular to the 3-fold axis directions of cube, similarly at $\theta_{\rm c1}$ as noted above.  Then, the solid angle of the 8 triangular surfaces on icosahedron changes from $\frac{\Omega_{ijk}}{4\pi}=\frac{1}{4}$ to $-\frac{1}{4}$ as $\theta$ increases from $\theta_{\rm c3}^{-}$ to $\theta_{\rm c3}^{+}$. Hence, the change in the topological charge is $(-\frac{1}{4}-\frac{1}{4})\times 8=-4$ so that the topological charge $n=+3$ for $\theta_{\rm c2}<\theta<\theta_{\rm c3}$ changes as  $n=3-4=-1$ for $\theta>\theta_{\rm c3}$.  

As for the anti-whirling state characterized by $n=-3$ for $\theta_{{\rm c}2}<\theta<\theta_{{\rm c}3}$, the topological charge changes as $n=+1$ for $\theta>\theta_{{\rm c}3}$. 

The absolute value of the topological charge $|n|$ is plotted in the ground-state phase diagram Fig.~\ref{fig:PD}. As shown in 
%Fig.~\ref{fig:or_sq}, 
Fig.~\ref{fig:AF}(b) 
the whirling state and the anti-whirling state both appear in the bcc unit cell in the region denoted by the filled orange-square symbols in Fig.~\ref{fig:PD}, where the topological charge $n$ explained above and $-n$ both appear on the corner and center icosahedrons, respectively. 
In Figs~\ref{fig:PD}(a) and \ref{fig:PD}(b), the vertical dotted lines are plotted at $\theta=\theta_{\rm c1}$, $\theta_{\rm c2}$, and $\theta_{\rm c3}$, respectively.

%------------------------------------------------------------------
\section{Summary and discussion}
\label{sec:SD}

%By performing numerically-exact calculation, 
In this study, we have performed complete enumeration of all configurations of magnetic moments in the 1/1 AC with the expanded unit cell with two icosahedrons under the periodic boundary condition. Then, 
we have determined the ground-state phase diagram of the uniaxial magnetic model with the AFM interactions in the 1/1 AC.
Depending on the ratio of the N.N. and N.N.N. interactions and the anisotropy angle $\theta$, various non-collinear magnetic states appear in the phase diagram, whose magnetic space groups are identified as eight-kinds: $Im'm'2$,  $Im'm'm'$, $I2'2'2$, $R\bar{3}'$, $R_{\rm R}\bar{3}$, $I_{\rm P}m'\bar{3}'$, $P2'/m'$, and $P\bar{1}$. 
For $J_2/J_1\ge 1$, the $\Gamma$-point antiferromagnetic orders with the magnetic propagation vector $(000)$ are stabilized, while for $J_2/J_1<1$, the antiferromagnetic orders with the magnetic propagation vector $(111)$ share the wide regions of the phase diagram. 
In the most areas of the phase diagram, the total magnetic moment per icosahedron or unit cell is zero, while the ferrimagnetic orders with the net ordered moments along the 2-fold axis direction of the bcc lattice appear for small $J_2/J_1$ and $0^{\circ}\le\theta\lsim 30^{\circ}$ regime. 
The cancellation of the total magnetic moment per icosahedron or per unit cell in the antiferromagnetic orders is due to antiferromagnetic interactions $J_1$ and $J_2$, while emergence of the ferrimagnetic order is non trivial, which is regarded as the consequence of geometrical frustration effect arising from the triangular geometry of the rare-earth sites on the icosahedron.  

We have shown that the topological magnetic textures characterized by the topological charge $n$ on the icosahedron appear in the ground-state phase diagram with $|n|=1$, 2, and 3.  
We have calculated the total chirality ${\bm \chi}^{\rm T}$ for each magnetic state, which plays a role as a fictitious emergent magnetic field. Then, we have found that the net ${\bm \chi}^{\rm T}$ is non-zero for the ferrimagnetic states denoted by the pink-circle and the black-triangle symbols, which appear for small $J_2/J_1$ and $0^{\circ}\le\theta\lsim 20^{\circ}$ in the phase diagram.   
%Both phases are monoclinic magnetic space group and 
Then, the topological Hall effect is expected to appear in the Hall conductivity.  
%$\sigma_{\mu\nu}$ perpendicular to the principal axis of the monoclinic phase. 

In the vortex order, the alignment of swirling magnetic moments is realized on the icosahedron, which is characterized by the toroidal magnetic moment ${\bm T}_{\rm IC}$ along the principal axis of $R\bar{3}'$. Since the cluster magnetic toroidal moments are aligned in the ferroic manner in the 1/1 AC, the electromagnetic effect where the magnetization is induced by applying electric field in the perpendicular direction ${\bm M}\perp {\bm E}$ is expected to appear. 

We have also clarified the degeneracy of each ground state, which is expected to be reflected in the number of the magnetic domains. This is useful to analyze the experimental observations considering the effect of domains, which inevitably exists in real materials.

In the present study, 
%we have performed numerical calculations in the 1/1 AC with the expanded unit cell with two icosahedrons under the periodic boundary condition [see the frame box in Fig.~\ref{fig:lattice}(b)]. 
we have determined the ground state by calculating the energies of all the eigenstates of the uniaxial model (\ref{eq:H}). It is, however, noted that 
there is a possibility that the magnetically ordered state(s) with larger unit cell or incommensurate ordered state is stabilized as the ground state. 
Although the magnetic phases sharing wide region in Fig.~\ref{fig:PD} are expected not to be altered by performing such calculations, in the narrow phase or around the border of the two phases new (incommensurate) phase(s) may appear. In any case, such a calculation is left for future studies. 

Recently, the CEF in the Ce-based QC and AC Au-SM-Ce has been analyzed theoretically~\cite{WFK2026}, where the magnetic easy axis has been shown to change drastically along three different directions in Fig. 1(f) depending on the ratio of the valences of screened ligand ions. Namely, uniaxial anisotropy is almost along the pseudo 5-fold-axis [$z$ axis in Fig.~\ref{fig:Tb_local}(f)] direction, almost perpendicular direction from the pseudo 5-fold-axis within the mirror $(yz)$ plane, or the perpendicular direction to the mirror plane. In the ternary Au-SM-Ce compounds, there exists mixed sites of Au and SM. Then, the mixed sites colored in green shown in Fig.~\ref{fig:Tb_local}(f) is occupied by Au or SM stochastically. Hence, the uniaxial direction is considered to change site by site depending on the configuration of the mixed sites. It is speculated that this can be the origin of the emergence of the spin glass behavior observed in 1/1 AC Au-SM-Ce at low temperatures~\cite{Imura,Sawano,Muro}.

So far, experimental identification of the magnetic structures of the rare-earth based 1/1 ACs have been reported for Cd-Tb~\cite{Das} and Au-SM-R (R=Tb, Ho)~\cite{Sato2019,Hiroto,Nawa2023,Labib2023} and the latter has  positive Curie-Weiss temperatures indicating the FM interactions. The present study has revealed theoretically that rich non-collinear ordered phases are stabilized for the AFM interactions in the 1/1 AC. Hence, it is interesting to explore the magnetic structures shown in this study in the rare-earth based 1/1 ACs with negative Curie-Weiss temperatures, which is left for future subject. 
Future studies such as neutron and resonant-X-ray measurements are highly desired. 
%To identify the magnetic structures, our results on the degeneracy of each ground state are expected to be useful to analyze the effect of domains as noted above.  

%
% Each of the commands below will create an unnumbered section with the appropriate heading.
% Remove any sections that are not relevant for your article.
% All sections except suppdata will be removed if the [anonymous] option is used.
% See iopjournal-guidelines.pdf for more information.
%

\ack{The author acknowledges K. Momma for expanding functionality of VESTA to draw the magnetic moments as well as atoms~\cite{Momma}. 
%This work was supported by JSPS KAKENHI Grant Numbers JP18K03542, JP19H00648, JP22H04597, JP22H01170, JP23K17672, and JP24H01675.
}

\funding{This work was supported by JSPS KAKENHI Grant Numbers JP18K03542, JP19H00648, JP22H04597, JP22H01170, JP23K17672, and JP24H01675}
% This section is a list of funder names and grant numbers

\roles{S.W. conceived the study and led the project. Theoretical calculation was performed by S.W. The manuscript was written by S.W.}
% List author names and the contributions made to the article, using terms from the NISO Contributor Roles Taxonomy (CRediT) https://credit.niso.org

%\data{Sample text inserted for demonstration.}
% For more information on IOP Publishing's research data policy see: https://publishingsupport.iopscience.iop.org/questions/research-data/

%\suppdata{Sample text inserted for demonstration.}

%\section*{References}

\end{document}